\documentclass[reprint,aps,prb,twocloumn,superscriptaddress,showpacs]{revtex4-2}
\usepackage{float}
\usepackage{graphicx}
\usepackage{color}
\usepackage{amsmath}
\usepackage{amsfonts}
\usepackage{amssymb}
\usepackage{framed}
\usepackage{float}
\usepackage[normalem]{ulem}
\usepackage[dvipsnames]{xcolor}
\usepackage{soul}
\sethlcolor{green}
\usepackage[colorlinks=true,citecolor=blue,linkcolor=blue]{hyperref}
\usepackage{parskip}
\def\be{\begin{equation}}
\def\ee{\end{equation}}

\def\bi{\begin{itemize}}
\def\ei{\end{itemize}}
\def\bn{\begin{enumerate}}
\def\en{\end{enumerate}}
\def\bea{\begin{eqnarray}}
\def\eea{\end{eqnarray}}

\def\ba{\begin{array}}
\def\ea{\end{array}}
\def\bd{\begin{displaymath}}
\def\ed{\end{displaymath}}

\usepackage{lineno}

\begin{document}

\title{Electron viscosity and device-dependent variability in four-probe electrical transport in ultra-clean graphene field-effect transistors}

\author{Richa P. Madhogaria}
\thanks{equal contribution}
\email{richapm@iisc.ac.in}
\affiliation{Department of Physics, Indian Institute of Science, Bangalore 560012, India}
\author{Aniket Majumdar}
\thanks{equal contribution}
\email{aniketm@iisc.ac.in}
\affiliation{Department of Physics, Indian Institute of Science, Bangalore 560012, India}   
\author{Nishant Dahma}
\thanks{equal contribution}
\email{nishantdahma@iisc.ac.in}
\affiliation{Department of Physics, Indian Institute of Science, Bangalore 560012, India}
\author{Pritam Pal}
\affiliation{Department of Physics, Indian Institute of Science, Bangalore 560012, India}
\author{Rishabh Hangal}
\affiliation{Department of Physics, Indian Institute of Science, Bangalore 560012, India}
\author{Kenji Watanabe}
\affiliation{Research Center for Electronic and Optical Materials, National Institute for Materials Science, 1-1 Namiki, Tsukuba 305-0044, Japan}
\author{Takashi Taniguchi}
\affiliation{Research Center for Materials Nanoarchitectonics, National Institute for Materials Science, 1-1 Namiki, Tsukuba 305-0044, Japan}
\author{Arindam Ghosh}
\email{arindam@iisc.ac.in}
\affiliation{Department of Physics, Indian Institute of Science, Bangalore 560012, India}
\affiliation{Center for Nano Science and Engineering, Indian Institute of Science, Bangalore 560012, India}


\begin{abstract}{
Hydrodynamic electrons in high-mobility graphene devices have demonstrated great potential in establishing an electronic analogue of relativistic quantum fluid in solid-state systems. One of the key requirements for observing viscous electron flow in an electronic channel is a large momentum-relaxation path, a process primarily limited by electron-impurity/phonon scattering in graphene. Over the past decade, multiple complex device geometries have been successfully employed to suppress momentum-relaxing scattering mechanisms; however, experimental observations have been found to be sensitive to the device fabrication process and architecture, raising questions about the signature of electron hydrodynamics itself. Here, we present a study on multiple ultra-clean graphene field-effect transistors (FETs) in a simple, rectangular four-terminal device architecture. Using electrical transport measurements, we have characterised the pristine quality of the graphene FETs and examined the variation of electrical resistance in the doped regime as a function of carrier density and temperature. Our results reveal strong device-dependent variability even in the most simple architecture that we attribute to competing momentum-conserving and momentum-relaxing scattering mechanisms, as well as coupling to contacts. Further, we have proposed a phenomenological method for analysing the results, which yields transport parameters in accordance with recent experiments. This simple experimental strategy and analysis can serve as an efficient tool for extracting the viscous electronic contribution in state-of-the-art high-mobility graphene FETs.
}
\end{abstract}

\maketitle
\section{Introduction \label{intro}}

The hydrodynamic flow of electrons \cite{gurzhi1968hydrodynamic} in two-dimensional (2D) mesoscopic channels has revealed non-trivial quantum phases of matter, leading to electronic and thermal transport signatures in Poiseuille flow \cite{sulpizio2019visualizing, ku2020imaging}, negative local resistance \cite{bandurin2016negative}, Wiedemann-Franz (WF) law violation \cite{crossno2016observation, majumdar2025universality}, superballistic conduction \cite{krishna2017superballistic}, Mott formula violation \cite{ghahari2016enhanced}, Gurzhi effect \cite{huang2023electronic, gurzhi1963minimum}, electron vorticity \cite{palm2024observation, aharon2022direct}, Planckian scattering \cite{gallagher2019quantum} and giant thermal diffusivity \cite{block2021observation}. This fluid-like motion of electrons has recently garnered significant attention, thanks to the emergence of new material platforms, such as high-mobility 2D layered semiconductors \cite{dean2010boron, lin2019towards}, which enable momentum-conserving electronic interactions to dominate electron-impurity scattering. Monolayer graphene, mechanically exfoliated from naturally occurring graphite, is one such material that has sparked widespread interest in the field of electron hydrodynamics due to its low intrinsic charge inhomogeneity ($n_\mathrm{min}(0)$) and linear electronic band-structure, causing the conduction and valence bands to meet at the Dirac point (DP) \cite{castro2009electronic}. In the low-carrier density regime, particularly near DP, electrons and holes undergo Planckian scattering \cite{planckianscattering}, thereby giving rise to a relativistic fluid of massless Dirac fermions \cite{muller2009graphene} - also referred to as the Dirac fluid \cite{crossno2016observation, ku2020imaging, gallagher2019quantum}. On the other hand, in doped graphene, momentum-conserving electron-electron interactions lead to the formation of a viscous charged fluid \cite{bulkandshear}. This firmly establishes ultra-clean graphene as a promising platform for demonstrating electron hydrodynamics in a solid-state condensed matter system.

\begin{figure*}[tbh]
    \centering
    \includegraphics[width=1.0\linewidth]{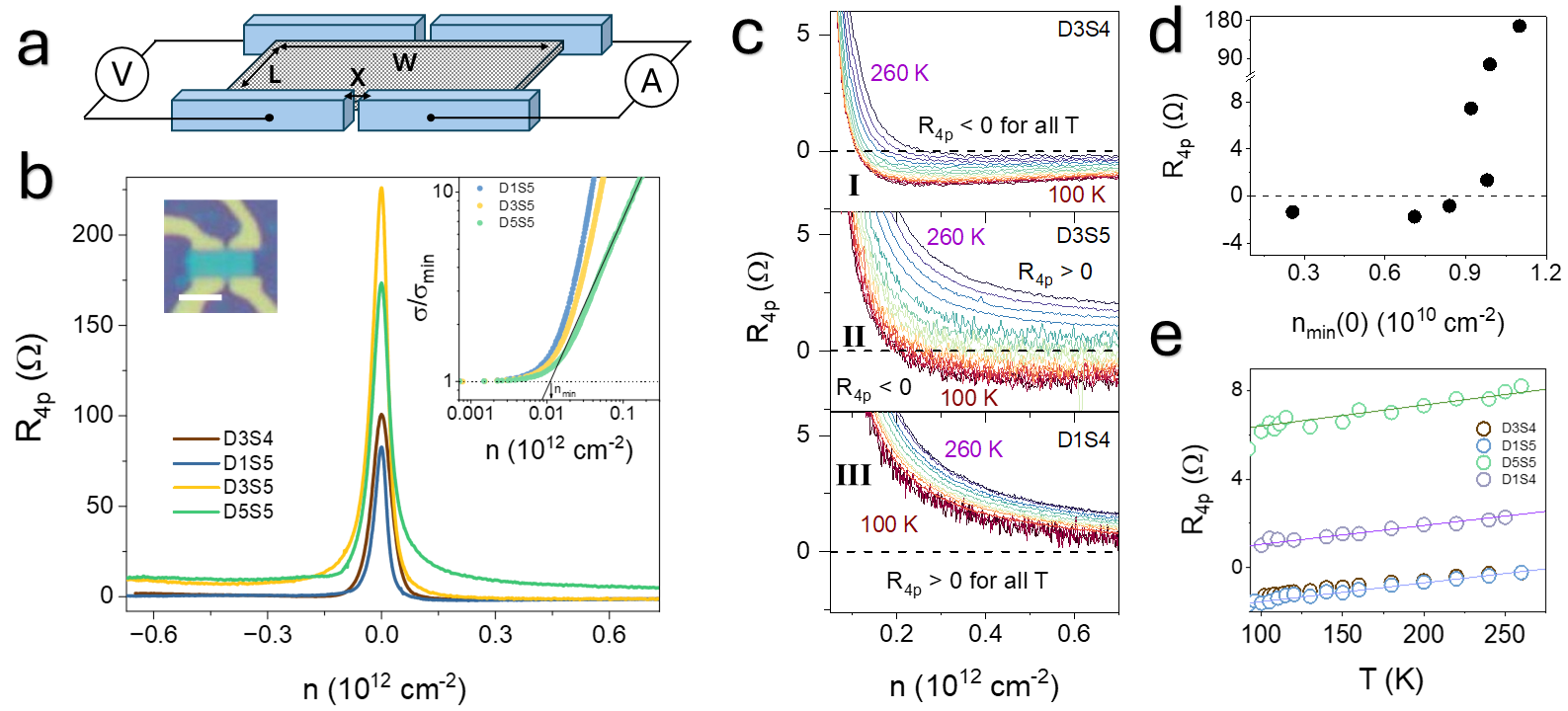}
    \caption{\textbf{Four-probe electrical resistance in an ultra-clean graphene channel:} (a). Schematic of the hBN-encapsulated graphene (grey) device consisting of four electrical contacts (blue). (b) $R_\mathrm{4p}$ vs carrier density $n$ for devices D3S4, D1S5, D3S5 and D5S5 at $T = 100$~K. Left inset - Optical micrograph of a device. Scale-bar along both axes is $3$~$\mu$m. Right inset - Normalized electrical conductivity ($\sigma/\sigma_\mathrm{min}$) vs $n$ for devices D1S5, D3S5 and D5S5 at $T = 100$~K. The horizontal dashed line indicates $\sigma = \sigma_\mathrm{min}$, and the solid line indicates $\sigma \propto \sqrt{n}$. The intersection of these two lines has been used to calculate the total charge inhomogeneity $n_\mathrm{min}$. (c) $R_\mathrm{4p}$ vs $n$ for devices D3S4 (class-I), D3S5 (class-II), and D1S4 (class-III) as a function of $T$ at intermediate and high densities. The horizontal dashed line indicates $R_\mathrm{4p} = 0$~$\Omega$. (d) $R_\mathrm{4p}$ vs $n_\mathrm{min}(0)$ at $T = 100$~K and $n = 4\times10^{11}$~cm$^{-2}$. The horizontal dashed line indicates $R_\mathrm{4p} = 0$~$\Omega$. (e) $R_\mathrm{4p}$ vs $T$ for devices D3S4, D1S5, D5S5 and D1S4 at $n = 5\times10^{11}$~cm$^{-2}$.}
    \label{fig1}
\end{figure*}

Despite their cleanliness, graphene-based van der Waals (vdW) heterostructures patterned in conventional device architectures — such as Hall bars and interdigitated contact geometries — often yield ambiguous signatures of hydrodynamic flow due to the non-locality of transport or complex trajectories associated with ballistic and thermal electronic motion. To address this issue, studies employing advanced nanofabrication techniques have investigated novel device layouts such as constrictions \cite{krishna2017superballistic}, Corbino disks \cite{talanov2024observation, zeng2024quantitative}, vicinity geometries \cite{bandurin2016negative}, anti-dot superlattices \cite{estrada2025superballistic}, crenellations \cite{geometriccontrol}, and clover-leaf structures \cite{aharon2022direct}. These architectures, although efficient in detecting the response of electron-electron scattering processes,  have yielded conflicting results, with some experiments, using a vicinity geometry, reporting negative local resistance \cite{bandurin2016negative, bandurin2018fluidity}, while others, using alternative contact geometries \cite{zeng2024quantitative, talanov2024observation, ponomarenko2024extreme}, failing to observe the same. Theoretical studies are also in direct conflict regarding these experimental findings being signatures of viscous electron flow \cite{whirlpool, shytov2018particle}. In this context, distinguishing genuine signatures of hydrodynamic transport from artefacts induced by complex device architecture becomes important. A unified framework to account for device-to-device variations in transport data is currently lacking, further complicating the correct interpretation of existing experimental results.

In this work, we aim to resolve this issue by developing a strategy to address device-to-device variability in four-probe resistance measurements ($R_\mathrm{4p}$) in high-mobility graphene. Our approach involves the use of high-quality graphene-based vdW heterostructures exhibiting electron mobilities ranging between $10^5-10^6$~cm$^2$~V$^{-1}$~s$^{-1}$ at temperatures as high as $\approx 240$~K (See Table~\ref{table1}) and mean free paths ($l_\mathrm{mfp}$) greater than $1$~$\mu$m at all temperatures. We carried out four-terminal electrical transport measurements (schematic in Fig.~\ref{fig1}a) on several devices as a function of gate voltage ($V_\mathrm{G}$) and temperature ($T$), observing both positive and negative resistance regimes. To analyse the gate voltage dependence of $R_\mathrm{4p}$, we employed a heuristic formalism based on the assumption that our device architecture has competing contributions from electron hydrodynamics and non-local coupling between current and voltage contacts. The resulting analysis offers a phenomenological platform for analysing the device-to-device variability observed in electrical transport measurements in mesoscopic systems with few or no scattering centres. Our findings yield realistic values of parameters for hydrodynamic electron flow in graphene and reveal its sensitivity to intrinsic device inhomogeneities and the specifics of contact configuration.

\begin{table*}[ht] \label{table1}
\centering
\caption{ \textbf{List of samples:} Details specifying the channel dimensions ($L$ and $W$), the separation between adjacent contacts ($X$), aspect ratio $L/W$, intrinsic charge inhomogeneity ($n_\mathrm{min}(0)$) and differential electron mobility ($\mu$) calculated at $n=10^{11}~\mathrm{cm}^{-2}$ and $T = 240~\mathrm{K}$, of the samples used in the main text.}
 
 \begin{tabular}{|| c | c | c | c | c | c | c ||} 
 \hline
 Device & $L$ ($\mu$m) & $W$ ($\mu$m) & $X$ ($\mu$m) & $L/W$ & $n_\mathrm{min}(0) (10^{10}$~cm$^{-2}$) & $\mu~(10^5\mathrm{cm^2/V.s})$ \\ [0.75ex] 
 \hline\hline 
 D2S1 & 2.1 & 4.2 & 0.42 & 0.5 & 1.48 & 2.2\\ 
 D1S4 & 1 & 6.7 & 0.36 & 0.15 & 0.98 & 5.7\\
 D3S4 & 1.5 & 5.5 & 0.44 & 0.27 & 0.25 & 18.0\\
 D4S4 & 0.6 & 0.8 & 0.48 & 0.75 & 1.02 & 0.35\\ 
 D1S5 & 1.4 & 8.2 & 0.85 & 0.17 & 0.64 & 51.3\\
 D3S5 & 1.1 & 4.6 & 0.72 & 0.23 & 0.84 & 8.4\\
 D5S5 & 1.2 & 5.7 & 0.83 & 0.21 & 0.92 & 3.9\\
 D1S6 & 1.4 & 3.4 & 0.5 & 0.41 & 0.99 & 1.3\\
 D4S6 & 1.1 & 2.4 & 0.5 & 0.45 & 1.1 & 1.1\\ 
 D4S9 & 1.1 & 12 & 0.92 & 0.09 & 0.25 & 10.1\\ 
 D6S9 & 1.3 & 8.1 & 0.96 & 0.16 & 0.3 & 11.5\\ 
 D1S10 & 0.9 & 4.5 & 0.39 & 0.2 & 1.1 & 1.5\\  [1ex]
 \hline
 \end{tabular}
\end{table*}

\section{Device configuration}
The array of devices chosen for this work involved monolayer graphene sandwiched between multiple layers of hexagonal boron nitride (hBN) dielectric. The Raman spectra of the graphene flakes were carefully monitored to confirm the pristine nature of graphene. The thickness of the hBN flakes varied between $20$--$30$~nm to ensure robust encapsulation of the graphene layer, thereby preventing atmospheric exposure. The surface roughness of hBN was kept below $2$~nm to minimise the effect of strain and contamination from organic residues and other short-range Coulomb scatterers (Details in Supplementary Section S1). The heterostructure was assembled using a dry transfer technique and eventually patterned into rectangles with four edge-contacted gold electrodes (Fig.~\ref{fig1}b - Left Inset). Although the length $L$ of the devices was kept approximately constant ($1-1.5$~$\mu$m), the width ($W$) of the channel was allowed to vary from $\approx 1$~$\mu$m to $10$~$\mu$m across different devices in order to examine the dependence of the electrical conductivity on the device dimensions (Table~\ref{table1}). We ensured that the $R_\mathrm{4p}$ measurements in the chosen device geometry comprised mostly of the local electrical response \cite{abanin2011} of the channel by keeping the separation between the current injection contacts and the voltage probes, labelled using $X$ (Fig.~\ref{fig1}a), significantly smaller than the length of the channel and the width of the contacts. This facilitated minimization of the classical contribution to any possible nonlocal response (See Supplementary Section S2). Additionally, $X$ was varied between $0.3$~$\mu$m and $1.0$~$\mu$m (confirmed by atomic force microscopy) to study the contribution of any non-local coupling between the current and voltage contacts and other electrostatic boundary conditions (See Supplementary Section S3 for additional details on these boundary conditions).
  
\begin{figure*}[tbh]
    \centering
    \includegraphics[width=0.7\linewidth]{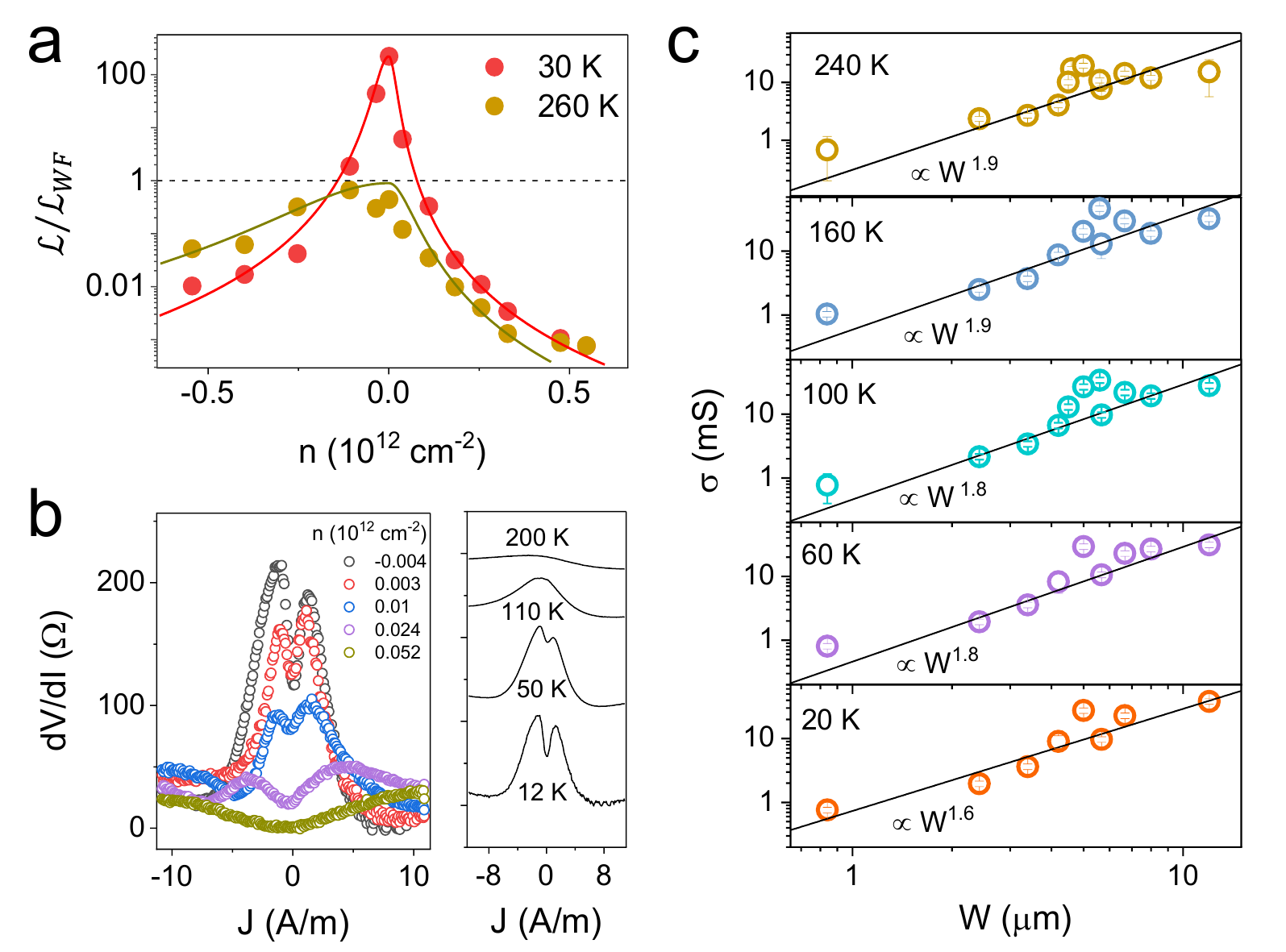}
    \caption{\textbf{Evidence of viscous electron flow:} (a) Violation of WF Law - Plot of $\mathcal{L}/\mathcal{L}_\mathrm{WF}$ as a function of $n$ for D2S1 at two different temperatures. (b) Current-induced decrease in differential resistance - [Left Panel] $dV/dI$ vs $J$ in D3S5 for different carrier densities near DP. [Right Panel] $dV/dI$ vs $J$ for $n = 10^{10}$~cm$^{-2}$ stacked across different temperatures. (c) Width dependence of electrical conductivity - Plot of $\sigma$ vs $W$ across $5$ different temperatures for $n = 10^{11}$~cm$^{-2}$. The solid line serves as a guide to the eye and is proportional to $W^\beta$, $\beta \simeq 1.8 \pm 0.1$.}
    \label{fig2}
\end{figure*}

\section{Experimental Results}
The transfer characteristics of some of the measured devices at $100$~K, obtained using the measurement schematic illustrated in Fig.~\ref{fig1}a, are shown in Fig.~\ref{fig1}b. The devices exhibited Lorentzian-type dependence of $R_\mathrm{4p}$ on the carrier density $n$, with a sharp increase near DP and an asymptotic saturation to near-zero values at large densities. The high quality of the channel was attributed to its low $n_\mathrm{min}(0)$. The plot of $\sigma(n, T)/\sigma_\mathrm{min}(T)$ as a function of $n$ (Fig.~\ref{fig1}b - Right Inset) has been utilised to evaluate the total charge inhomogeneity $n_\mathrm{min}(T)$ at temperature $T$, where $\sigma(n, T) = L/R_\mathrm{4p}W$ is the electrical conductivity at temperature $T$ and carrier density $n$, and $\sigma_\mathrm{min}(T)$ is the minimum electrical conductivity at temperature $T$. Since $n_\mathrm{min}(T) = n_\mathrm{min}(0) + n_\mathrm{th}(T)$ where $n_\mathrm{th}(T) = (2\pi^3/3)(k_\mathrm{B}T/h v_\mathrm{F})^2$ refers to the thermally excited carrier density at temperature $T$ \cite{xin2023giant, ponomarenko2024extreme, majumdar2025universality} ($h$, $k_\mathrm{B}$ and $v_\mathrm{F}$ are the Planck's constant, Boltzmann's constant and Fermi velocity respectively), $n_\mathrm{min}(0)$ can be evaluated from the plot of $n_\mathrm{min}(T)$ vs $T$ (Details in Supplementary Section S4). For all the measured devices, $n_\mathrm{min}(0) \lesssim 10^{10}$~cm$^{-2}$ (see Table~\ref{table1}). 

The experimentally observed $n$-dependence of $R_\mathrm{4p}$ (in the electron-doped regime) across different devices, as indicated in Fig.~\ref{fig1}c, can be broadly categorised into three distinct classes - (I) \textbf{Top Panel} (device D3S4) : For $n \gtrsim 10^{11}$~cm$^{-2}$, $R_\mathrm{4p} < 0$ for the entire range of temperature under consideration ($100$~K $\le T \le 260$~K). Once $R_\mathrm{4p}$ becomes negative, it keeps decreasing with increasing $n$ till it reaches a minimum after which $R_\mathrm{4p}$ starts increasing weakly and approaches zero at high $n$. (II) \textbf{Middle Panel} (device D3S5): For $n \gtrsim 10^{11}$~cm$^{-2}$, $R_\mathrm{4p} < 0$ at low $T$ and becomes positive above some characteristic device-dependent temperature scale. Even in such cases, once $R_\mathrm{4p}$ falls below zero at an intermediate $n$, it no longer crosses over to $R_\mathrm{4p} > 0$ at higher values of $n$ at the same $T$. (III) \textbf{Bottom Panel} (device D1S4): For $n \gtrsim 10^{11}$~cm$^{-2}$, $R_\mathrm{4p} > 0$ for the entire range of experimentally accessible $T$. These observations constitute the full extent of device-to-device variation that can occur in electrical transport measurements in ultra-clean graphene FETs. We also note that, due to pronounced electron-hole asymmetry in the devices, the hole-doped regime often exhibits a complicated $n$-dependence of $R_\mathrm{4p}$ \cite{xia2011origins} and does not yield observations similar to the electron-doped regime. 

The tendency of $R_\mathrm{4p}$ to assume a positive or negative value at high $n$ has been found to have a non-trivial dependence on $n_\mathrm{min}(0)$ (Fig.~\ref{fig1}d). This is illustrated for $n \sim 5 \times 10^{11}$~cm$^{-2}$ and $T=100$~K, where we observe that for low $n_\mathrm{min}(0)$, $R_\mathrm{4p}$ fluctuates around negative values and starts increasing for $n_\mathrm{min}(0) \gtrsim 10^{10}$~cm$^{-2}$, eventually crossing over to positive resistances. Also, $R_\mathrm{4p}$ for $n \gtrsim 10^{11}$~cm$^{-2}$ across all the devices increases linearly with $T$ (Fig.~\ref{fig1}e). This can be attributed to an electron-phonon scattering mediated $R_\mathrm{4p} = R_0 + \gamma T$ where $R_0$ is a function of $n$ and $\gamma = (\pi^2 D^2k_\mathrm{B}L)/(2e^2h\rho_\mathrm{m}v_\mathrm{s}^2 v_\mathrm{F}^2 W)$ with $D$, $e$, $\rho_\mathrm{m}$ and $v_\mathrm{s}$ respectively being the acoustic deformation potential, electronic charge, mass density and the speed of sound in graphene \cite{kaasbjerg2012unraveling, efetov2010controlling, hwang2008acoustic}. Using $\rho_\mathrm{m} = 7.6 \times 10^{-7}$~kg~m$^{-2}$ and $v_\mathrm{s} = 2.6 \times 10^{4}$~m~s$^{-1}$ \cite{efetov2010controlling}, the fitted lines for different devices in Fig.~\ref{fig1}e correspond to $D = 14-17$~eV, which is in good agreement with previous experimental reports \cite{bolotin2008temperature, dean2010boron, chen2008intrinsic}.

\begin{figure*}[tbh]
    \centering
    \includegraphics[width=1.0\linewidth]{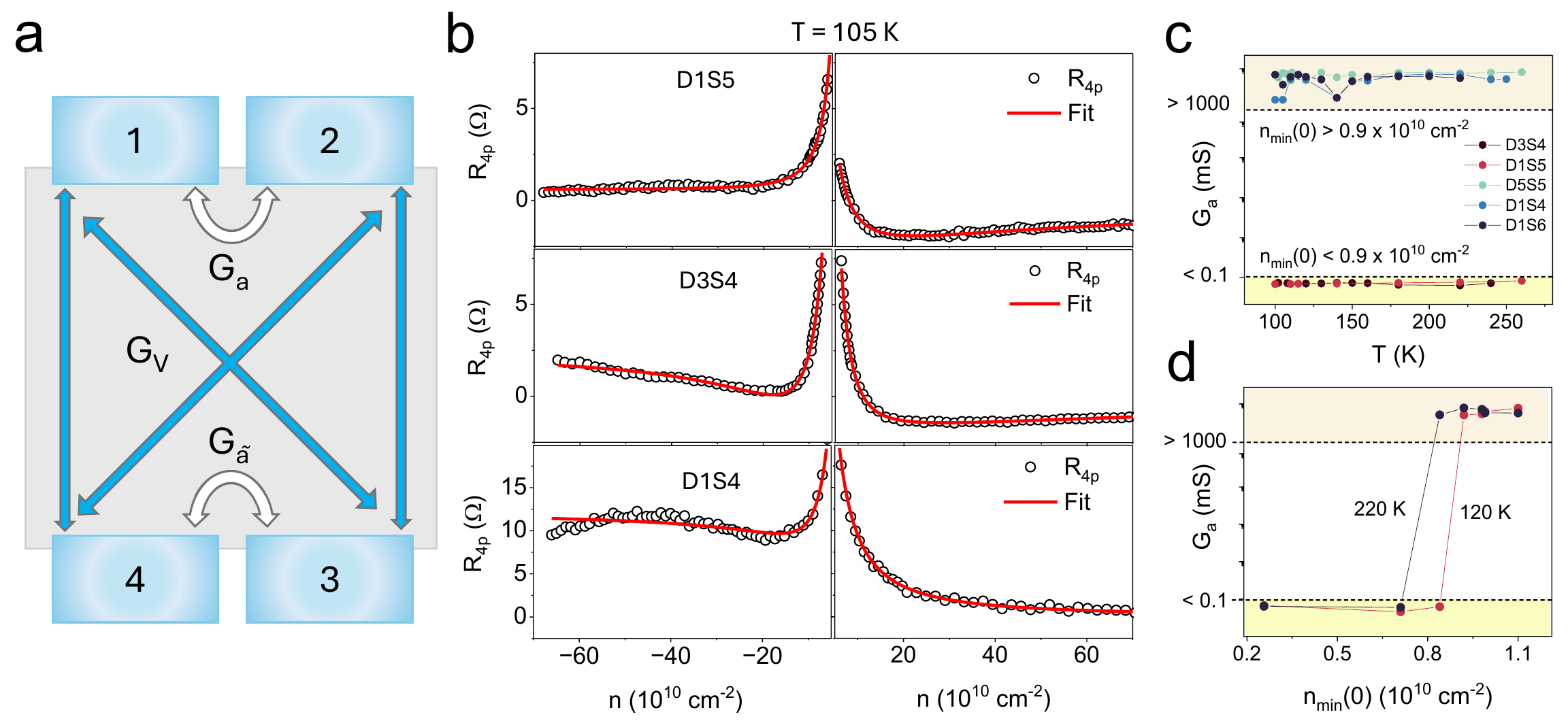}
    \caption{\textbf{The phenomenological model:}  (a) Schematic of different conductances used in the model. The grey area corresponds to the channel, connected to four blue-coloured electrical contacts ($1,~2,~3$ and $4$). Blue arrows represent $G_\mathrm{V}$ while white arrows represent $G_\mathrm{a},~G_\mathrm{\tilde{a}}$. (b) Experimentally obtained $R_\mathrm{4p}$ (black circles) as a function of $n$ for $\lvert n \rvert \ge 5 \times 10^{10}$~cm$^{-2}$ and $T = 105$~K for devices D1S5, D3S4 and D1S4, along with the best-fitted curves (red solid lines) as per our formalism. Left Panels - hole-doped, Right Panels - electron-doped. (c)  $G_\mathrm{a}$ versus $T$ for different devices. (d) $G_\mathrm{a}$ versus $n_\mathrm{min}(0)$ for different $T$. The coloured bars in figures (c) and (d) indicate the range over which the fitted $G_\mathrm{a}$ varies. The yellow (beige) bar represents $G_\mathrm{a} < 0.1~(>1000)$ mS.}
    \label{fig3}
\end{figure*}

\section{Discussion}

\subsection{Viscous electron flow in ultra-clean graphene} 

Despite the variability in the transport data, the low $n_\mathrm{min}(0)$ of the channels can be associated with the dominance of electron-electron scattering. However, the device architecture plays a crucial role in exploring these viscous interactions - the most important criterion being the development of no-slip boundary conditions \cite{torre2015nonlocal} along the channel edges (See Supplementary Section S5). In addition to this, numerical simulations of the electric potential $\phi$ and the current density profile $J_y(x)$ in our sample geometry reveal that the onset of Poiseuille electron flow depends on a range of device-specific parameters (See Supplementary Section S6). This gives rise to a small parameter phase space, suitable for the development of viscous electronic flow, which can be easily affected by non-zero slip and trace amounts of electronic disorder (See Supplementary Section S7). 

From an experimental viewpoint, negative resistance has been attributed to electron viscosity in the past \cite{bandurin2016negative, bandurin2018fluidity}; however, alternative mechanisms have also been found to yield negative $R_\mathrm{4p}$ \cite{shytov2018particle}. As a result, the observation of $R_\mathrm{4p} < 0$ in some of our devices (Classes I and II) cannot be considered conclusive evidence of hydrodynamic electron flow. To resolve this, multiple experiments have been conducted - the first one involved thermal transport measurements using Johnson-Nyquist noise thermometry \cite{crossno2016observation, majumdar2025universality, gugnani2025dynamically}. Our measurements revealed a giant violation of the WF law across a range of carrier densities and temperatures, with the effective Lorenz number, defined as $\mathcal{L} = \kappa_\mathrm{e}/\sigma T$ ($\kappa_\mathrm{e}$ being the electronic component of thermal conductivity), overshooting the semiclassically expected value ($\mathcal{L}_\mathrm{WF} = \pi^2 k_\mathrm{B}^2/3e^2$) by almost two orders of magnitude,  as shown in Fig.~\ref{fig2}a (Details in Supplementary Section S8). The departure from the semiclassical WF law strongly suggests the decoupling of charge and heat flow pathways in the presence of strong momentum-conserving scattering and the subsequent emergence of collective excitations, giving rise to an electron fluid in graphene \cite{principi2015}. In addition to this, we also observe a non-monotonic variation of the differential resistance $dV/dI$ as a function of injected current density $J$ -- another signature of viscous electron flow \cite{molenkapp}. This peculiar behaviour can be understood as follows: with increasing $J$, the electron-electron scattering rate increases, implying an increase in the number of momentum-conserving electron-electron collisions. Since these collisions are dissipationless, this eventually leads to a Knudsen ($dV/dI \propto J$) to Gurzhi ($dV/dI \propto 1/J$) crossover, as $l_\mathrm{ee}$ is varied from $>W$ to $<W$ (Fig.~\ref{fig2}b, red data points). As shown in Fig.~\ref{fig2}b, this effect sustains across a range of densities [Left Panel] and temperatures [Right Panel]. Further, we have investigated the dependence of $\sigma$ on the channel width $W$ (Fig.~\ref{fig2}c) and observed that $\sigma \propto W^\beta$, where $\beta \simeq 1.8 \pm 0.1$ for $T \ge 100$~K, which is very close to the quadratic dependence of conductivity on channel width - a key evidence of Poiseuille electronic flow \cite{levitov2016electron} (Details in Supplementary Section S9). Another key feature is the saturation of the electrical conductance $G$ in Class III devices to $G_\mathrm{ball}$, the Landauer-Sharvin resistance of a ballistic conductor \cite{landauerResistance, sharvinResistance}, for $l_\mathrm{ee} \ge W$ ($l_\mathrm{ee} = \hbar^2v_\mathrm{F}^2 k_\mathrm{F}/\alpha^2 k_\mathrm{B}^2 T^2$ being the electron-electron scattering length with $\alpha \approx 0.5$ the fine structure constant of graphene \cite{lucas2018hydrodynamics}) and a sharp decrease in $G/G_\mathrm{ball}$ to values lower than $1$ as $l_\mathrm{ee}/W$ decreases below $1$ \cite{majumdar2025universality} (See Supplementary Section S10). However, extracting electron viscosity in this device architecture is non-trivial due to the occurrence of negative electrical resistance in class-I and II devices.

\subsection{Phenomenological Analysis}

With transport signatures strongly suggesting viscous electron flow, we now investigate device-dependent variability in $R_\mathrm{4p}$ through a theoretical framework that captures the hydrodynamic conductivity in the form of a non-trivial transmission probability and an effective number of conducting modes (Details in Supplementary Section S11). Since we are interested in the DC electrical response, we divide the channel into electrical conductances $G_\mathrm{ij}$ (Fig.~\ref{fig3}a), existing between the i$^\mathrm{th}$ and j$^\mathrm{th}$ contacts. Transport within the bulk of our samples is governed by electron hydrodynamics as demonstrated earlier; hence, the relevant electrical conductance is the viscous conductance $G_\mathrm{V} = n^2e^2W^3/12\eta L$ \cite{torre2015nonlocal,lucas2018hydrodynamics} for a pristine channel exhibiting shear viscosity $\eta$. Further, since the detection of viscous electrons requires sub-micrometre spacing between the current and voltage contacts \cite{berdyugin2019measuring}, there is a non-local coupling between the current and voltage contacts along the graphene edge. The exact scattering mechanisms underlying this particular resistive contribution are unclear and have been accounted for by using the empirical coupling conductances $G_\mathrm{a}$ and $G_\mathrm{\tilde{a}}$ (Fig.~\ref{fig3}a). 
By simplifying this network of conductors (Details in Supplementary Section S12), the resulting four-probe resistance is evaluated as 

\begin{equation} \label{eqn}
    R_\mathrm{4p} = \frac{1}{8G_\mathrm{V}} \frac{G_\mathrm{a} G_{\tilde{\mathrm{a}}} - G_\mathrm{V}^2}{(G_\mathrm{a} + G_\mathrm{V})(G_{\tilde{\mathrm{a}}}+G_\mathrm{V})}
\end{equation}

where $G_{\tilde{\mathrm{a}}}$ and $G_\mathrm{a}$ represent $G_{12}$ and $G_{34}$ respectively, accounting for the asymmetry between the two components. The experimentally obtained $n$-dependence of $R_\mathrm{4p}$ was fitted using Eqn.~\ref{eqn} by considering $G_\mathrm{V}$, $G_\mathrm{a}$ and $G_{\tilde{\mathrm{a}}}$ to be of the mathematical form $G = b_1 n^{b_2}$, where $b_1$ and $b_2$ are fitting parameters (Details in Supplementary Section S13). We observe that the best fit is obtained by using an $n$-independent ($b_2=0$) $G_\mathrm{a}$. This reduces the total number of fitting parameters to four (See Supplementary Section S14), and the resulting best fit, as per Eqn.~\ref{eqn} has been highlighted in Fig.~\ref{fig3}b - the three panels show data at $105$~K from three devices of different $n_\mathrm{min}(0)$, two of which exhibit $R_\mathrm{4p} < 0$. Further, $G_\mathrm{a}$ exhibits a weak $T$ dependence for all the devices measured (Fig.~\ref{fig3}c), as expected from a tunnelling-mediated coupling process and has a strong dependence on $n_\mathrm{min}(0)$ (Fig.~\ref{fig3}d). The phenomenological coupling $G_\mathrm{a}$ can be attributed to a combination of scattering mechanisms, such as short-ranged electron-impurity scattering and diffusive scattering from the graphene boundary; however, the exact physical mechanism underlying $G_\mathrm{a}$ requires a detailed investigation. Equation~\ref{eqn} has been used to extract $G_\mathrm{V}$ and $G_\mathrm{a}$ from the $R_\mathrm{4p}$ versus $n$ data for all the devices. Our formalism describes the data accurately across the entire range of experimentally accessible temperatures (see fits at other temperatures in Supplementary Section S13). 

\begin{figure*}[tbh]
    \centering
    \includegraphics[width=1\linewidth]{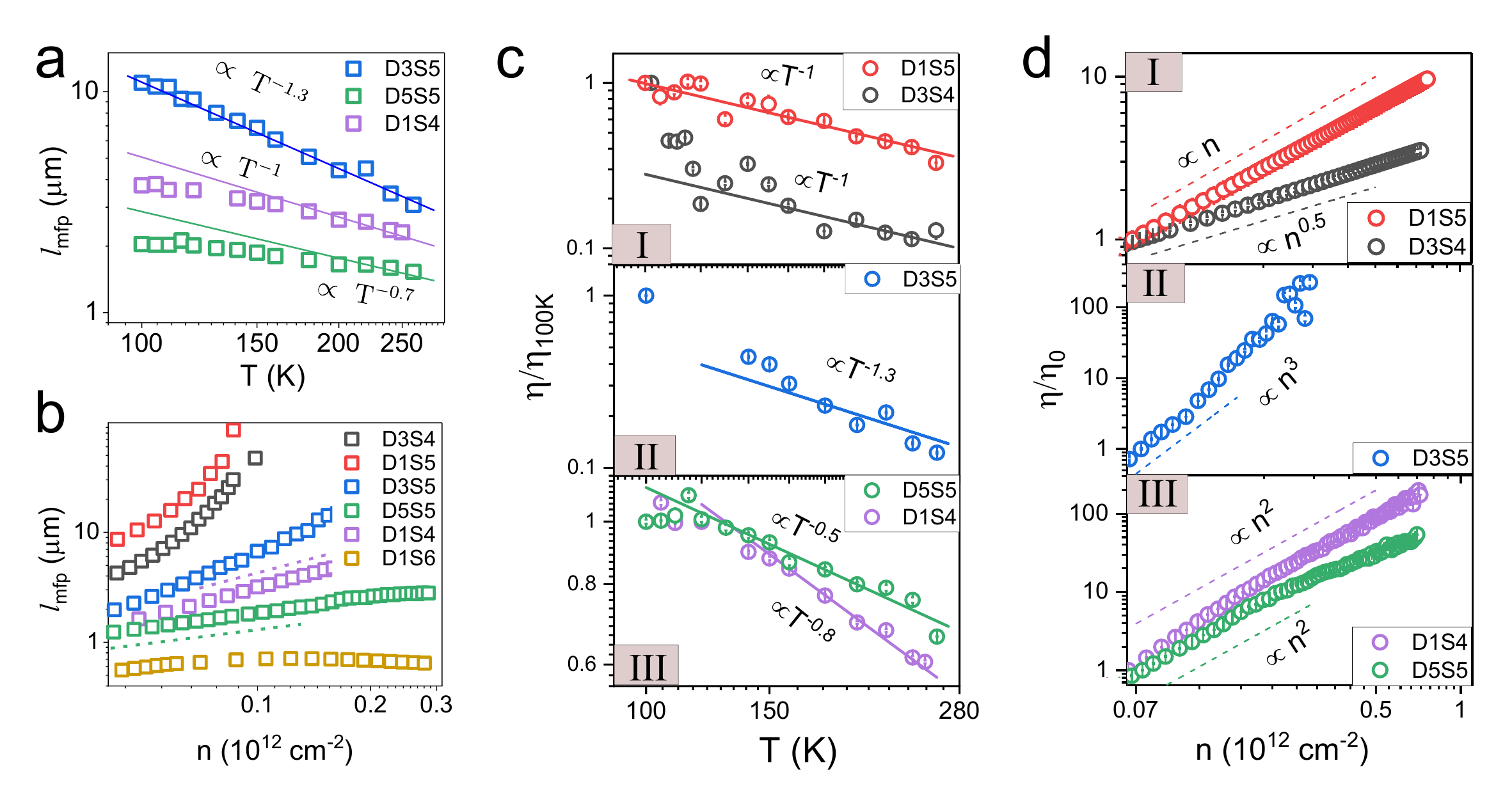}
    \caption{\textbf{Estimation of the electronic shear viscosity:} (a) Variation of $l_\mathrm{mfp}$ with $T$ for three devices, evaluated for $n = 10^{11}$~cm$^{-2}$. The solid lines represent $1/T^{0.7}$, $1/T$ and $1/T^{1.3}$ dependences. (b) $n$-dependence of $l_\mathrm{mfp}$ for four devices at $T = 105$~K. The purple and green dashed lines indicate $n$ and $\sqrt{n}$ dependence, respectively. (c) $\eta$ normalized with $\eta~(\mathrm{T=100~K})$ as a function of $T$ measured at $n = 10^{11}$~cm$^{-2}$. [Top to Bottom] Panel I: Devices D3S4 and D1S5 (Class-I), which display negative $R_\mathrm{4p}$ for $T \leq 260$~K. Panel II: Device D3S5 (Class-II), where $R_\mathrm{4p}$ makes a transition from negative to positive at $T \geq 160~\mathrm{K}$. Panel III: Devices D1S4 and D5S5 (Class-III) with positive $R_\mathrm{4p}$ throughout. (d) Plots of $\eta$ normalized with $\eta_0 = \eta~(n\ \mathrm{{=7\times10^{10}~cm^{-2}}})$ as a function of $n$ at $T = 150$~K. [Top to Bottom] The devices follow the same scheme as in (c).}
    \label{fig4}
\end{figure*}

\subsection{Temperature and density dependence of electronic viscosity in doped graphene}

We now utilise the outcome of the phenomenological analysis to evaluate key parameters of the viscous electron fluid and compare those with the theoretical expectation. In the electron-doped regime, Poiseuille flow requires $l_\mathrm{ee} \ll l_\mathrm{mr}$ (momentum relaxation length) to be the shortest length scale. The spatial extent of the viscous flow, however, depends on the viscous length \cite{huang2023electronic} which, in turn, is related to $l_\mathrm{mr} = \mathrm{min}(l_\mathrm{mfp},W)$, where $l_\mathrm{mfp}$ is evaluated using $\sigma = (2e^2/h)k_\mathrm{F}l_\mathrm{mfp}$ as per the Drude model. Based on this observation, viscous electrons can exist in a graphene FET under two scenarios - (a) $l_\mathrm{mr} \gg W, l_\mathrm{ee}$ (D3S4, D1S5) and (b) $l_\mathrm{ee} \ll l_\mathrm{mr} \lesssim W$ (D1S4, D3S5, D5S5). 

Irrespective of the varying $n_\mathrm{min}(0)$, all devices presented in this study unanimously exhibit $l_\mathrm{mfp}$ to be monotonically decreasing with $T$ (Fig.~\ref{fig4}a) such that $l_\mathrm{mfp}$ scales as $1/T^{\delta}$, where $\delta = 1.0 \pm 0.3$. This $1/T$ dependence can be attributed to electron-acoustic phonon scattering in monolayer graphene \cite{hwang2008acoustic, efetov2010controlling}. Additionally, $l_\mathrm{mfp}$ is found to scale as $n^\theta$ for many of the devices, as presented in Fig.~\ref{fig4}b, where $\theta \in [0.5,1]$ - this $n$-dependence might owe its origin to long-range electron-impurity scattering processes \cite{sarkar2015role}. In total, these observations indicate that even in our high-mobility graphene samples, electron-electron interactions in the doped regime compete with coexisting electron-phonon and electron-impurity scattering processes for $T \ge 100$~K. As a result, the magnitude of $l_\mathrm{mr}$ becomes a key factor in determining $\eta$.

We also observe that $l_\mathrm{mfp} \gg l_\mathrm{ee}$, suggesting $\tau^{-1}_\mathrm{mfp} \ll \tau^{-1}_\mathrm{ee}$, in contrast to what is expected from Matthiessen's rule \cite{Hui2025}, where scattering rates from independent mechanisms add up. This discrepancy can be explained by the fact that scattering mechanisms in a viscous electron fluid do not originate from independent sources and can be described using the addition of electrical conductances – the anti-Matthiessen’s rule \cite{guo2017higher}, which states that $\tau_\mathrm{mfp} = \tau_\mathrm{mr} + \tau_\mathrm{ee}$ (Details in Supplementary Section S15). For devices with $l_\mathrm{ee} \ll l_\mathrm{mr} \lesssim W$, namely D3S5, D5S5 and D1S4 (see Table~\ref{table1}), the momentum relaxation occurs at $l_\mathrm{mr} \simeq l_\mathrm{mfp} \lesssim W$. The resulting $\eta$, evaluated using $\eta = n^2e^2Wl_\mathrm{mfp}^2/12LG_\mathrm{V}$ \cite{lucas2016transport}, is plotted in Fig.~\ref{fig4}c-II and III as a function of $T$. On the other hand, for Class-I devices like D3S4 and D1S5, where $l_\mathrm{mfp} \gg W$ (see Table~\ref{table1}) indicates $l_\mathrm{mr} = W$, the viscous length scale is determined by $W$. The shear viscosity, in such devices, is given by $\eta = n^2e^2W^3/12LG_\mathrm{V}$ and depicted in Fig.~\ref{fig4}c-I as a function of $T$. We observe that all devices, irrespective of the class, exhibit $\sim 1/T$ dependence over the range of experimentally accessible temperatures, particularly for $T \geq 150$~K, with the exponent varying around $-1.0 \pm 0.2$.  This $\eta \propto 1/T$ observation is in sharp contrast to the Fermi liquid (FL) prediction \cite{steinberg1958viscosity, bulkandshear, muller2009graphene, yudhistira2025nonmonotonic}, which states that 
\begin{equation} \label{etaeqn}
    \eta \propto \dfrac{E_\mathrm{F}^4}{\hbar v_\mathrm{F}^2 (k_\mathrm{B}T)^2} \propto \dfrac{n^2}{T^2}
\end{equation}
However, the $1/T$ dependence of $\eta$ observed in our devices is not an anomaly; rather, similar results have been reported in previous experimental works \cite{sarypov2025temperature, geometriccontrol, krishna2017superballistic, berdyugin2019measuring, zeng2024quantitative}. Although this observation lacks a rigorous theoretical description, this can be attributed to the temperature smearing of the Fermi surface \cite{sarypov2025temperature, geometriccontrol} at higher temperatures where $T \approx T_\mathrm{F}$ and consequently Eqn.~\ref{etaeqn} is no longer valid.

Further, the $n$-dependence of $\eta$ shows a non-trivial variation across the devices measured (Fig.~\ref{fig4}d). The majority of the measured devices, belonging to Classes II and III, indicate a quadratic dependence of $\eta$ on $n$ at intermediate densities, as expected from FL theory \cite{bulkandshear, muller2009graphene, yudhistira2025nonmonotonic}. However, Class I devices like D1S5, D3S4 show $\eta \propto n$ and $\eta \propto n^{0.5}$ respectively at different density ranges, as demonstrated in Fig.~\ref{fig4}d-I, which do not have consistent theoretical explanations but are in accordance with a few of the previous experimental results in vicinity and Corbino disk geometries, where $\eta \propto n$ \cite{ku2020imaging, gugnani2025dynamically} and $\eta \propto \sqrt{n}$ \cite{talanov2024observation} have been reported respectively.

\section{Summary}

Our samples exhibit a range of $T$ and $n$-dependences for $\eta$ in the doped limit for the same device architecture, despite showing clear signatures of viscous electron flow, as seen in Figs.~\ref{fig2}a-c. The expected FL-like $\eta \propto n^2/T^2$ dependence is absent; rather, the experimentally determined $\eta$ depicts a different picture, with the scales varying from $\eta \propto \sqrt{n}/T$ to $n^2/T$. These peculiar $n$ and $T$-dependences of $\eta$ at high carrier densities cannot be explained using conventional theories of electron hydrodynamics. Recently, newer theories have been developed that consider a combination of hydrodynamic and ballistic transport, referred to as the tomographic regime of electronic transport \cite{hofmann2022collective,ledwith2019tomographic}. The theory also predicts the existence of scale-dependent conductance and viscosity in 2D electron fluids. In addition to this, it also supports the observation $\eta \propto n^{5/3}/T$ \cite{zeng2024quantitative}, which is very close to what has been observed in devices D1S5, D5S5 and D1S4 at intermediate and high densities. However, further theoretical investigation is needed for a deeper understanding.
 
\par In conclusion, we have presented a simple strategy to address the device-dependent variability in electrical transport in extremely high-quality graphene devices with an effective model configured using a simple rectangular geometry with four electrical leads. Our results provide a comprehensive avenue to compute the shear viscosity of the electronic fluid, even for systems exhibiting apparent negative resistance. Our findings can be utilised to realise the simultaneous existence of multiple electron flow patterns, ranging from Poiseuille flow to tomographic flow, within the same device architecture in state-of-the-art ultra-clean graphene devices.

\section{Acknowledgement}
The authors gratefully acknowledge the usage of the MNCF and NNFC facilities at CeNSE, IISc. The authors thank Subroto Mukerjee and Sriram Ramaswamy for insightful discussions. The authors also acknowledge the usage of DST-FIST-funded Oxford-WITec Alpha 300R Confocal Raman/PL Spectroscope, Central Instruments Facility, Department of Physics, IISc. A.G. acknowledges financial support from J. C. Bose Fellowship and a project under NanoMission, Department of Science and Technology, India. P.P. thanks the Ministry of Education, Govt. of India, for the Prime Minister's Research Fellowship (PMRF). R.H acknowledges financial support from the INSPIRE Fellowship, Department of Science \& Technology, Govt. of India. K.W. and T.T. acknowledge support from the JSPS KAKENHI (Grant Numbers 21H05233 and 23H02052), the CREST (JPMJCR24A5), JST and World Premier International Research Center Initiative (WPI), MEXT, Japan.

\bibliography{ref}

\end{document}


\title{Supplementary Information for Electron viscosity and device-dependent variability in four-probe electrical transport in ultra-clean graphene field-effect transistors}

\author{Richa P. Madhogaria}
\thanks{equal contribution}
\email{richapm@iisc.ac.in}
\affiliation{Department of Physics, Indian Institute of Science, Bangalore 560012, India}
\author{Aniket Majumdar}
\thanks{equal contribution}
\email{aniketm@iisc.ac.in}
\affiliation{Department of Physics, Indian Institute of Science, Bangalore 560012, India}   
\author{Nishant Dahma}
\thanks{equal contribution}
\email{nishantdahma@iisc.ac.in}
\affiliation{Department of Physics, Indian Institute of Science, Bangalore 560012, India}
\author{Pritam Pal}
\affiliation{Department of Physics, Indian Institute of Science, Bangalore 560012, India}
\author{Rishabh Hangal}
\affiliation{Department of Physics, Indian Institute of Science, Bangalore 560012, India}
\author{Kenji Watanabe}
\affiliation{Research Center for Electronic and Optical Materials, National Institute for Materials Science, 1-1 Namiki, Tsukuba 305-0044, Japan}
\author{Takashi Taniguchi}
\affiliation{Research Center for Materials Nanoarchitectonics, National Institute for Materials Science, 1-1 Namiki, Tsukuba 305-0044, Japan}
\author{Arindam Ghosh}
\email{arindam@iisc.ac.in}
\affiliation{Department of Physics, Indian Institute of Science, Bangalore 560012, India}
\affiliation{Center for Nano Science and Engineering, Indian Institute of Science, Bangalore 560012, India}
    
\maketitle

\renewcommand*\contentsname{Table of Contents}
\tableofcontents
\listoffigures
\newpage

\section{Device fabrication} \label{Fabrication}
Hexagonal boron nitride (hBN) and graphene flakes were mechanically exfoliated onto a \ce{Si^{++}/SiO2} substrate with a 285 nm thick oxide layer. To enhance the adhesion of monolayer graphene, the substrate was pre-treated with \ce{O2} plasma etching prior to exfoliation \cite{graphene-exfoliation-method}. Flake identification was initially performed using optical contrast relative to the substrate \cite{optical-contrast-paper}, followed by atomic force microscopy (for hBN) and Raman spectroscopy (for graphene) \cite{ferrari_graphene} to characterise the cleanliness of the device. 

\begin{figure}[tbh]
    \centering
    \includegraphics[width=1\linewidth]{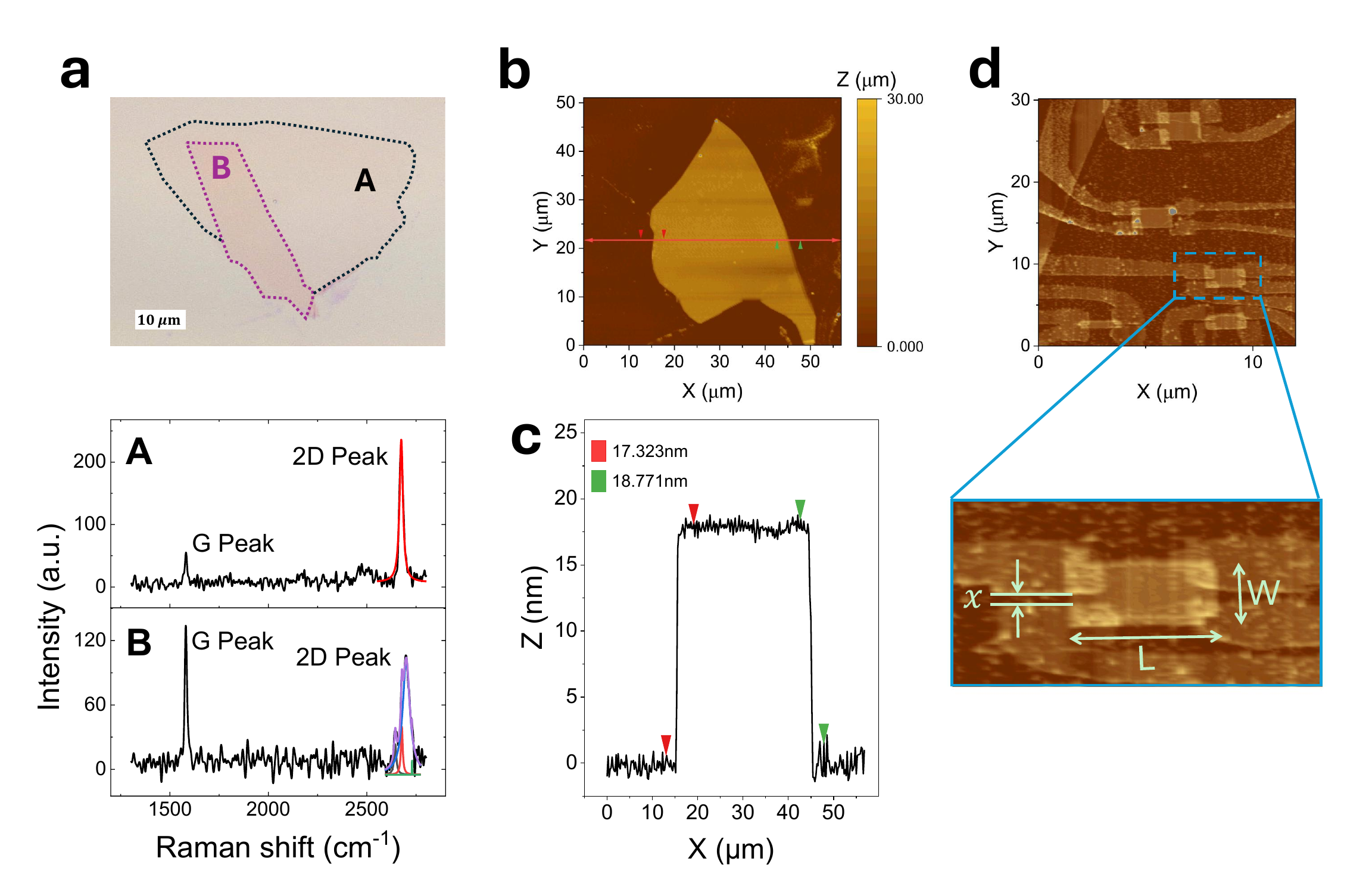}
    \caption[Characterization of van der Waals layers]{\textbf{Characterization of the van der Waals layers: } (a) [Top] An optical image of the graphene flake with regions A (in black) and B (in purple) marked with dotted lines. The scale bar provided with the image is $10$~$\mu$m. [Bottom] Raman spectra for regions A (black) and B (purple), respectively (given in the top image), from Raman shift $1300$~cm$^{-1}$ to $2800$~cm$^{-1}$. G and 2D peaks have been labelled along with fitted curves for 2D peaks.  (b) AFM image of an hBN flake. A line-scan marked with the red solid line is shown in (c). Red and green cursor pairs have been inserted to estimate the thickness and roughness of the flake. (c) Z-Height data of the line-scan in (b). The red and green cursor pairs correspond to heights $17.323$~nm and $18.771$~nm, respectively. (d) [Top] AFM image of multiple devices fabricated from a single stack. [Bottom] Magnified AFM image of a single device represented in blue. The top and bottom two contacts are current and voltage contacts, respectively. Dimensions of the device have been marked as $L$, $W$ and $X$ representing the length, width and separation distance between adjacent current and voltage contacts, respectively.}
    \label{fig:device-raman-afm}
\end{figure}

Fig. \ref{fig:device-raman-afm}a shows the Raman spectra of a graphene flake consisting of mono- and bilayer regions where G ($\sim1583$~cm$^{-1}$) and 2D ($\sim2699$~cm$^{-1}$) peaks have been labelled \cite{ferrari_graphene}. The G and 2D peaks arise from first and second-order inelastic scattering processes where the electronic momentum is relaxed with the help of phonons. The presented Raman spectra also show the absence of the D peak ($\sim1350$~cm$^{-1}$), which indicates a pristine graphene with no defects, edges or vacancies in the lattice \cite{clean_transfer_graphene}. The ratio of intensities of the 2D and G peaks ($I_\mathrm{2D}/I_\mathrm{G}$) is an indication of the doping concentration, with undoped pristine graphene displaying the largest $I_\mathrm{2D}/I_\mathrm{G}$ \cite{Das2008}. The number of layers is further confirmed by the minimum number of Lorentzians required to fit the 2D peak (one for mono- and four for bilayer), each Lorentzian corresponding to one of the possible scattering configurations. In addition to this, while the line-width of the Raman G-peak depends predominantly on phonon anharmonicity \cite{Phonon-anharmonicity_raman, Ferrari2013} and electron-phonon coupling \cite{Ferrari2013,el-ph_raman}, that of the 2D peak depends not only on the aforementioned, but also on other mechanisms such as electron-hole dispersion anisotropy and electronic momentum uncertainty originating from electronic scattering \cite{ferrari_graphene}. Since line-widths of both peaks increase with disorder \cite{g-peak-fwhm_raman, Das2008, Ferrari2013}, they can be considered as an important parameter for quantifying the intrinsic inhomogeneity in the graphene flake. The full width at half-maximum (FWHM) of the Raman G and 2D peaks presented in Fig. \ref{fig:device-raman-afm}a are $14.2$~cm$^{-1}$ and $20.8$~cm$^{-1}$ respectively, which can be considered sufficiently low. Such graphene flakes, characterised by a high $I_\mathrm{2D}/I_\mathrm{G}$ ratio, and sharp Raman peaks (low FWHM) were chosen for fabrication of the van der Waals (vdW) heterostructure.

During mechanical exfoliation, the topography of hBN flakes is often contaminated by tape residues and stress-induced cracks. These result in the creation of impurity scatterers at the graphene-hBN interface, which hampers the carrier mobility of the graphene channel \cite{hBN-blisters}. Using AFM characterisation, we have sought to identify clean, flat areas optimal for encapsulating an ultra-clean graphene monolayer. In Fig. \ref{fig:device-raman-afm}b, one such hBN flake has been shown, which exhibits a very clean and homogeneous top surface with minimal disorder and residue. A cross-sectional view of the flake, indicated by the red line in Fig. \ref{fig:device-raman-afm}b, shows a surface roughness of $\lesssim 0.7$~nm (Fig. \ref{fig:device-raman-afm}c), which can be considered as a key quantifier of interfacial cleanliness in graphene-hBN van der Waals heterostructures. 

After the desired flakes have been selected post-characterisation, a vdW-pickup-based dry-transfer technique is employed to assemble the hBN-encapsulated graphene stacks. In this technique, the desired top hBN flake present on the substrate is placed below the polydimethylsiloxane (PDMS)/polycarbonate (PC) stamp and brought into direct contact at a temperature $T \sim 110^\circ$~C. The monolayer graphene and bottom hBN flakes are subsequently picked up, thus creating a hBN/Gr/hBN van der Waals heterostructure. The encapsulated sample is finally brought into contact with a prepatterned Si$^{++}$/SiO$_2$ substrate and heated to $T \sim 180^\circ$~C until the film melts on the wafer. The melted PC film dissolves once the substrate containing the stack is immersed in chloroform. Finally, the transferred vdW assembly is cleaned using acetone and isopropyl alcohol (IPA) to remove any organic residue. Electron-beam lithography is used to pattern the rectangular channel and electrodes, followed by reactive-ion etching and thermal deposition of Cr ($6$~nm)/Au ($60$~nm). The surface topography of some of these devices, prepared from the same stack on a single Si$^{++}$/SiO$_2$ substrate, has been shown in Fig. \ref{fig:device-raman-afm}d. The magnified image of one of the devices depicts the AFM image of the contacts and the channel, and also indicates the minuscule spacing between the current and voltage probes. The Si substrate is also used as a back gate to tune the carrier density of the graphene channel. 

\section{Minimisation of nonlocal response}
The origin of the non-local electrical response in a mesoscopic sample can be attributed to classical effects (like development of electric potential outside the channel in accordance with Ohm’s law) as well as quantum effects (like band topology with non-zero Berry curvature, formation of ferroelectric domains, etc.). In this situation, we refer to the classical contribution, i.e., the non-local potential difference generated across the two voltage probes due to the ohmic voltage drop across the current probes. The classical contribution of the non-local response at distance $x$ from the current probe is given by $R_{NL}\sim\ \rho \exp[-\pi x/L]$ \cite{abanin2011}, where $\rho$ is the ohmic resistivity of the channel between the source and drain contacts, and $L$ is the separation between the voltage contacts. Hence, a small value of $x$ implies that the four-terminal measurement primarily reflects the local resistance, enabling the voltage probes to accurately capture the electrical response of the hydrodynamic electrons. In our devices, the value of $x$, which we can probe, is the separation between adjacent current and voltage contacts, which we call $X$. As shown in Fig. \ref{fig:Device-Schematic}, $X$ is usually kept much smaller than other physical dimensions of the device.

\section{Effect of electrostatic boundary conditions} 
\label{sec:electrostatic BC}
\begin{figure}[tbh]
    \centering
    \includegraphics[width=1\linewidth]{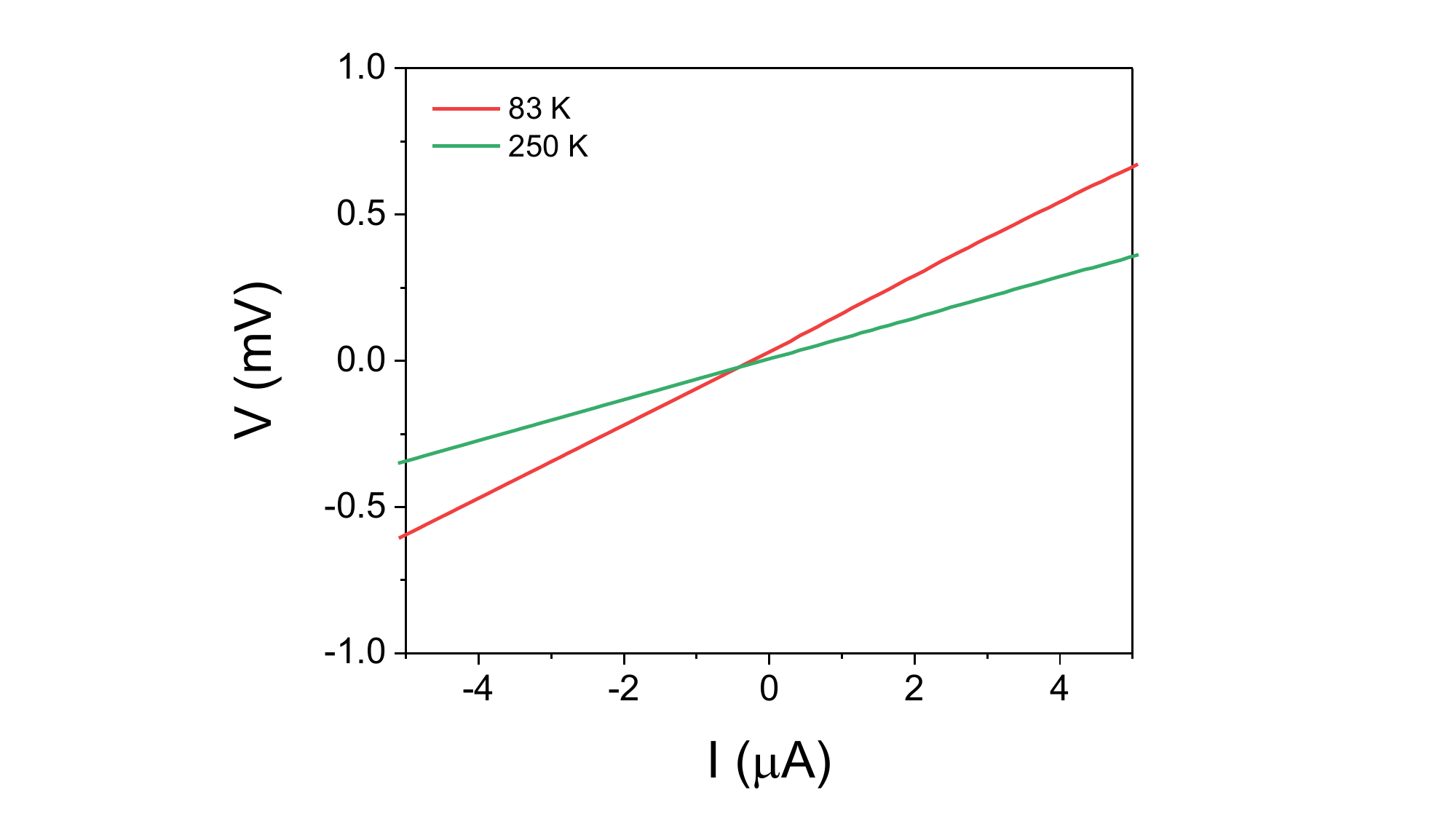}
    \caption[I-V characteristics of the device]{\textbf{I-V characteristics of the device: } $V-I$ curve for device D3S5 for temperatures $T = 83$~K (red) and $250$~K (blue). A linear trend indicates ohmic behavior and does not contribute to nontrivial electric field distribution in the graphene channel (see Section \ref{sec:electrostatic BC}). }
    \label{fig:I-V_characteristics}
\end{figure}

The metal electrodes used to probe the graphene device form one-dimensional edge contacts and are deposited using a combination of Cr and Au, which are well known for forming ohmic contacts at the metal-graphene interface \cite{wang2013one}. The I-V characteristics of our devices are shown in Fig. \ref{fig:I-V_characteristics}, which show linear ohmic behaviour at low electrical biases. This indicates that the electrical contacts are not responsible for the non-trivial electric field distribution within the channel. However, along the edges of the channel, due to the presence of electrostatically trapped charges along the edge between the current and voltage contacts, one cannot rule out the emergence of an offset in conductance, arising due to Fermi level pinning \cite{Moktadir2015}.  

\section{Calculations of intrinsic charge inhomogeneity and mobility}
\begin{figure}[tbh]
    \centering
    \includegraphics[width=1\linewidth]{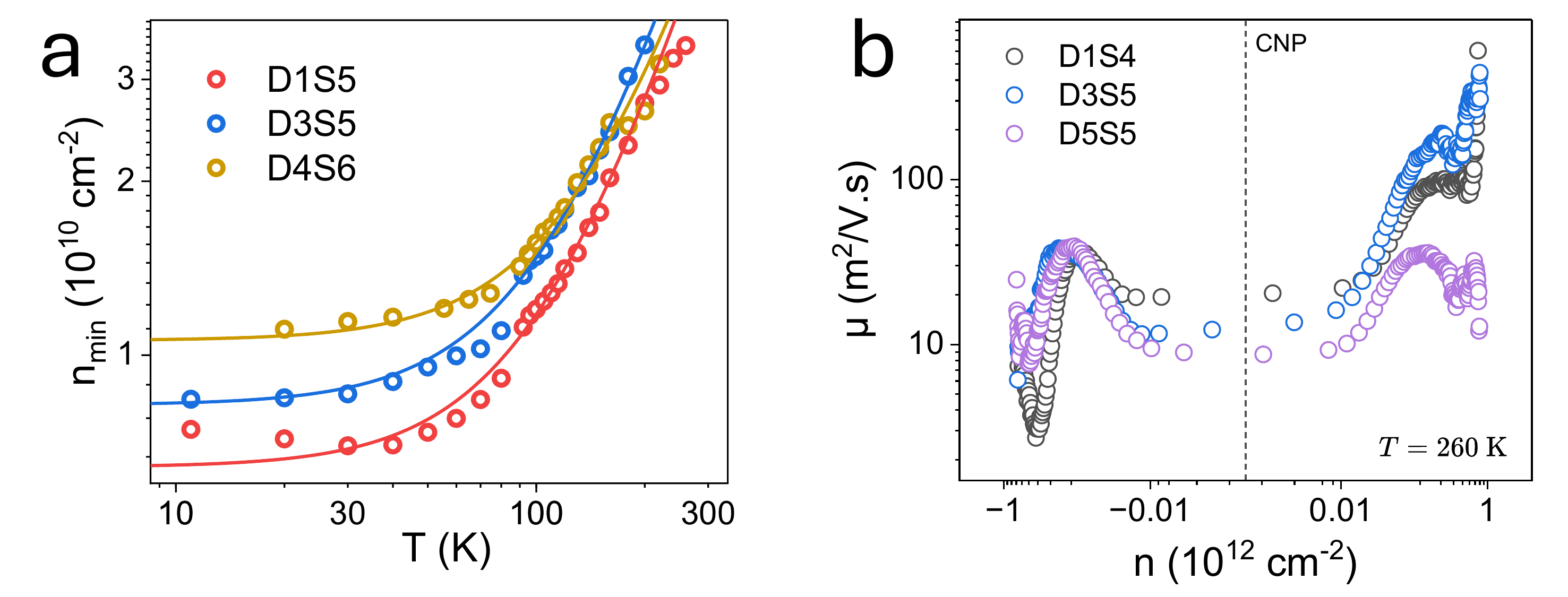}
    \caption[Electrical transport characteristics]{\textbf{Electrical transport characteristics:} (a) $n_{\mathrm{min}}$ vs $T$ ($10~\mathrm{K}\leq T \leq 260~\mathrm{K}$) for devices D1S5, D3S5 and D4S6 represented by circles of red, blue and yellow in colour respectively. The solid lines follow $n_\mathrm{min}(T) = n_\mathrm{min}(0) + \alpha \cdot n_\mathrm{th}(T)$ with $n_\mathrm{min}(0)$ as the only free parameter. (b) Electron mobility $\mu$ vs $n$ for three devices calculated at $T = 260~\mathrm{K}$ for $\lvert n \rvert \ge 5 \times 10^{9}$~cm$^{-2}$.}
    \label{fig:nmin}
\end{figure}

Charged impurities in the device significantly affect the electrical transport in the channel. This is reflected in the saturation of electrical conductivity $\sigma(n)$ towards $\sigma_\mathrm{min}$, as the Fermi energy is tuned to the Dirac point. The conductivity starts to increase as $n$ increases, and asymptotically tends towards $\sqrt{n}$ dependence. The carrier density at which $\sigma$ starts increasing drastically from $\sigma_\mathrm{min}$ is referred to as the total charge inhomogeneity $n_\mathrm{min}(T)$. It is evaluated by extrapolating the two curves $\sigma(n) = \sigma_\mathrm{min}$ and the best fit of $\sigma(n)$ at intermediate densities. The total charge inhomogeneity is a sum of two quantities - intrinsic charge inhomogeneity, $n_\mathrm{min}(0)$ and thermally excited carrier density, $n_\mathrm{th}(T) = (2\pi^3/3)(k_BT/hv_F)^2$. Fig. \ref{fig:nmin}a shows the values of $n_\mathrm{min}$ obtained for devices D1S5, D3S5 and D4S6 as a function of temperature. Furthermore, the temperature dependence of $n_\mathrm{min}(T)$ is explained by $n_\mathrm{min}(T) = n_\mathrm{min}(0) + \alpha \cdot n_\mathrm{th}(T)$ (the pre-factor $\alpha\simeq 0.6 - 0.8$ is likely due to the residual charge inhomogeneity) \cite{ponomarenko2024extreme}. $n_\mathrm{min}(0)$ is estimated from the $T$ dependence by noting the saturation of $n_\mathrm{min}(T)$ at lower temperatures ($T\le 100$~K). Another useful parameter to characterise device cleanliness is the carrier mobility ($\mu$), which is derived using $\mu = (1/e)(d\sigma/dn)$. Fig. \ref{fig:nmin}b shows the variation in carrier mobility with carrier density for three devices. Furthermore, electron mobility for devices with negative $R_\mathrm{4P}$ was estimated using the model described in \cite{ponomarenko2024extreme}.

\section{Validity of no-slip boundary conditions }
In order to estimate the viscous parameters in our devices, we assume the viscous conductance to be of the form $\sigma = n^2e^2l^2/12\eta$, where $n$ is the number density of the carriers, $e$ is the electronic charge, $l$ is the relevant length scale of interest and $\eta$ is the shear viscosity of the electron fluid. This formula inherently assumes no-slip boundary conditions, which imply that the electron fluid experiences sufficient friction at the device boundaries that the velocity is zero. A crucial step in fabricating our devices within the prescribed geometry involves reactive-ion etching (RIE), a dry etch technique that utilises reactive plasma and ion bombardment to ablate the patterned material. This process inevitably affects graphene edges, leading to increased microscopic roughness, lattice reconstruction near the edges, and the binding of foreign molecules, among other effects. Even under hBN encapsulation, these defects form, and the graphene lattice loses translational symmetry at the edges. 
Consequently, during the flow of current, charge localisation (which has been observed in graphene nanoribbons \cite{graphene-edge-importance}) along these etched boundaries causes electrons to have practically zero slip along the direction of current flow. From previous experimental reports on exfoliated bilayer graphene \cite{localization-length}, we can estimate an upper bound for the localisation length $\xi=500~\mathrm{nm}$. This suggests a boundary layer of localised charges, which is in good agreement with the widths of the devices used in this experiment. Graphene is known to exhibit three main edge types: zigzag, armchair, and reczag. After the inductively-coupled plasma treatment in RIE, graphene edges are not expected to be properly ordered, but rather segments of armchair/zigzag edges are most likely to form \cite{graphene-edges-reconstructions}. Numerical simulations show that with increasing defect density, conductance along these edges decreases significantly \cite{graphene-edges-simulation} and more so when molecular adsorption at the edges is taken into account.
However, the no-slip boundary condition can be relaxed by defining the parameter $l_b$ known as the boundary slip-length (details in Section \ref{sec:numerical-simulations}). This quantifies the tangential frictional force the electron fluid experiences due to the rough boundaries. 

\subsection{Evaluation of edge roughness using Raman spectroscopy}
\begin{figure}[tbh]
    \centering
    \includegraphics[width=1\linewidth]{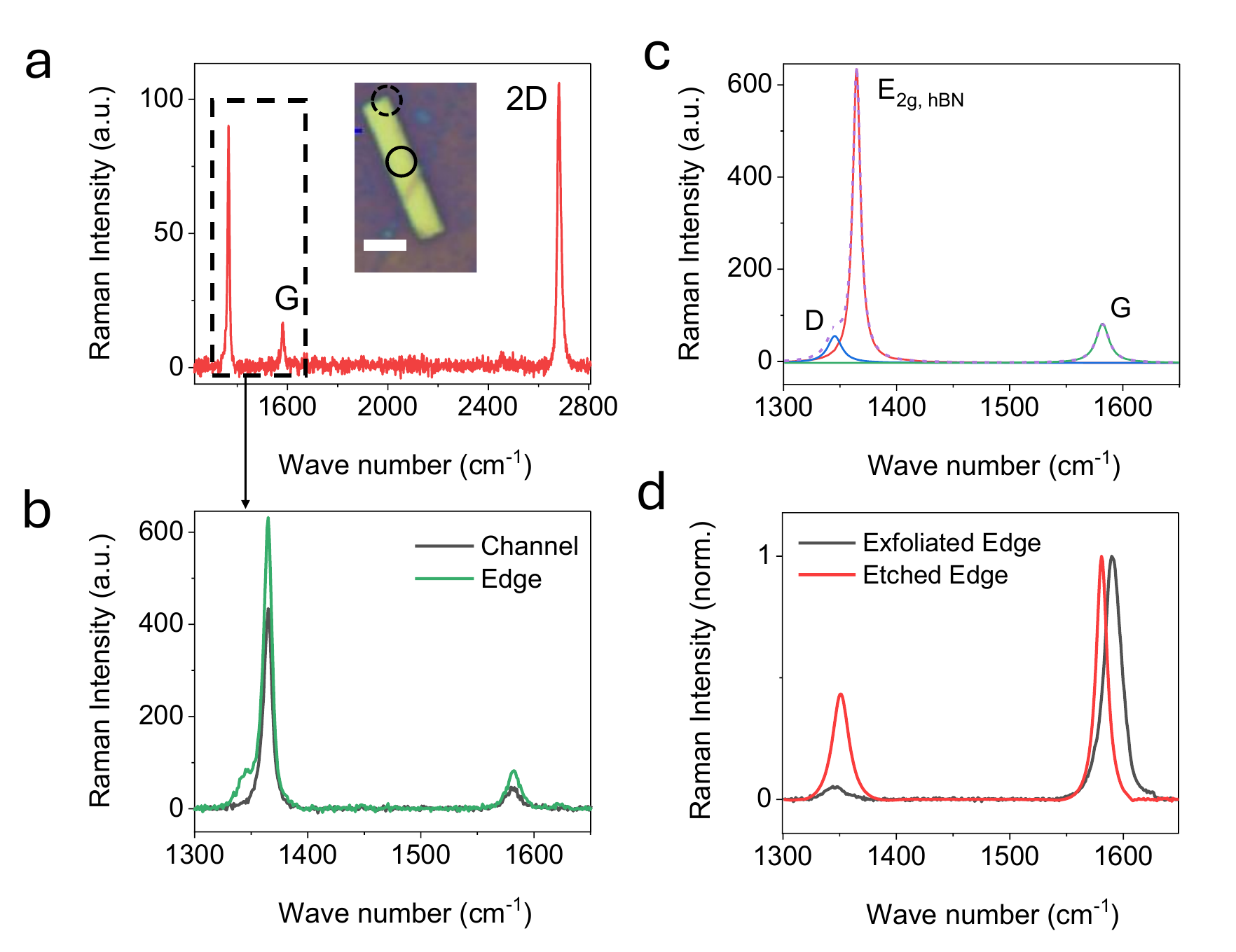} 
    \caption[Raman spectral analysis of the edge roughness of graphene]{\textbf{Raman spectral analysis of the edge roughness of graphene: } (a) Raman spectrum obtained at the edge (dashed circle) of the heterostructure. The spectrum was obtained using a $532$~nm laser with $8$~mW output power. An optical image of the heterostructure has been presented with its edge and channel areas marked with dashed and solid circles, respectively. The dashed rectangle representing the Raman spectrum from wavenumber $1300\rightarrow1650$~cm$^{-1}$ is zoomed in and presented in plot (b). The white scalebar is $5~\mathrm{\mu m}$. (b) Raman spectrum obtained at the channel (in black) and edge (in green) areas for the same parameters as (a). (c) Lorentzian fit of $D$, $E_\mathrm{2g}$(hBN) and $G$ peaks in the Raman spectrum, obtained in (b). The centres of the $D$, $E_\mathrm{2g}$(hBN) and $G$ peaks are $1345,~1364,$~and~$1582$~cm$^{-1}$ with their FWHMs being $15,~8$~and~$14$~cm$^{-1}$ respectively. (d) Variation of the Raman spectrum obtained at the edge of the van der Waals heterostructure to that obtained at the edge of an exfoliated graphene flake for the same incident laser power. 
    \label{fig:Raman spectrum}}
\end{figure}

A crude estimate of the edge roughness can be obtained by performing Raman spectroscopy along the RIE-treated edges of our hBN-encapsulated devices, as presented in Fig. \ref{fig:Raman spectrum}. The presence of a sharp $G$ peak at $\sim 1582$~cm$^{-1}$ and the absence of the $D'$ peak in the Raman spectrum recorded at the edge of one of the hBN-encapsulated devices (dashed circle) in Fig. \ref{fig:Raman spectrum}a indicates the average inter-defect distance $L_\mathrm{D}\gtrsim10$~nm \cite{Cancado2011}. Fig. \ref{fig:Raman spectrum}b compares the Raman spectra obtained at the edge (dashed circle) and the central channel (solid circle), where the $D$ peak (which is usually Raman forbidden but is visible near the edge due to disorder, broken translational symmetry and lattice defects) clearly appears for the edge spectra. Due to the $E_{2g}$ peak of hBN \cite{hBN-Raman} observed at $1364$~cm$^{-1}$, the graphene $D$ peak appears as a shoulder at $1345$~cm$^{-1}$. Furthermore, Raman spectroscopy was performed on the edges of exfoliated monolayer graphene at the same incident power in order to compare with the Raman spectra of RIE-treated graphene edge (see Fig.~\ref{fig:Raman spectrum}d). In order to quantify $L_\mathrm{D}$ and the defect density $n_\mathrm{D}$, the $D$, $E_{2g}$ and $G$ peaks were identified by performing a Lorentzian fit and subsequently obtaining the intensity ratios $I_\mathrm{D}/I_\mathrm{G} = 0.689$ for the RIE-treated edge and $I_\mathrm{D}/I_\mathrm{G} = 0.0489$ for the exfoliated graphene edge. The following relations were used to estimate $L_\mathrm{D}$ and $n_\mathrm{D}$ \cite{Cancado2011,Ferrari2013}:

\begin{gather}
    n_\mathrm{D}~(\mathrm{cm}^{-2}) = \frac{(1.8\pm0.5)\times10^{22}}{\lambda_L^4}\cdot\left(\frac{I_\mathrm{D}}{I_\mathrm{G}}\right) \label{eqn:nD}\\
    L_\mathrm{D}^2~(\mathrm{nm}^{2}) = (1.8\pm0.5)\times10^{-9} \lambda_L^4 \cdot\left(\frac{I_\mathrm{D}}{I_\mathrm{G}}\right)^{-1} \label{eqn:LD}
\end{gather}

Here, $\lambda_\mathrm{L}$ is the laser wavelength in nanometers ($532$~nm for the spectrometer used in our case). Using Eqs. \ref{eqn:nD} and \ref{eqn:LD}, for the RIE-treated edge and the exfoliated graphene edge, we obtain defect densities $n_\mathrm{D}^{\mathrm{RIE}}=(1.5 \pm 0.4) \times 10^{11}~$cm$^{-2}$ and $n_\mathrm{D}^{\mathrm{exf.}}=(1.1 \pm 0.3) \times 10^{10}~$cm$^{-2}$ respectively. The corresponding average inter-defect distances are therefore $L_\mathrm{D}^\mathrm{RIE}=14.4\pm7.6$~nm and $L_\mathrm{D}^\mathrm{exf.}=54.2\pm28.6$~nm respectively. This result implies that the RIE-treated edge is $\sim4\times$ rougher than the edge of exfoliated graphene.
The length scale $L_\mathrm{D}^\mathrm{RIE}\sim14$~nm is much smaller when compared to the physical dimensions of our RIE-treated devices, which are of the order of microns. Furthermore, boundary slip-length ($l_b$) quantifies how strongly the edges relax momentum of the electron fluid and is a dynamic quantity that depends on the inter-defect distance ($L_\mathrm{D}$), which is a static quantity. Although these quantities are inherently different, we expect them to be of the same order of magnitude. Thus, one can crudely expect $l_b$ to be of the order of $\sim$~$\mathcal{O}(10)$~nm, which supports our use of no-slip boundary conditions.

\section{Numerical simulations of viscous transport} \label{sec:numerical-simulations}
\begin{figure}[tbh]
    \centering
    \includegraphics[width=0.6\linewidth]{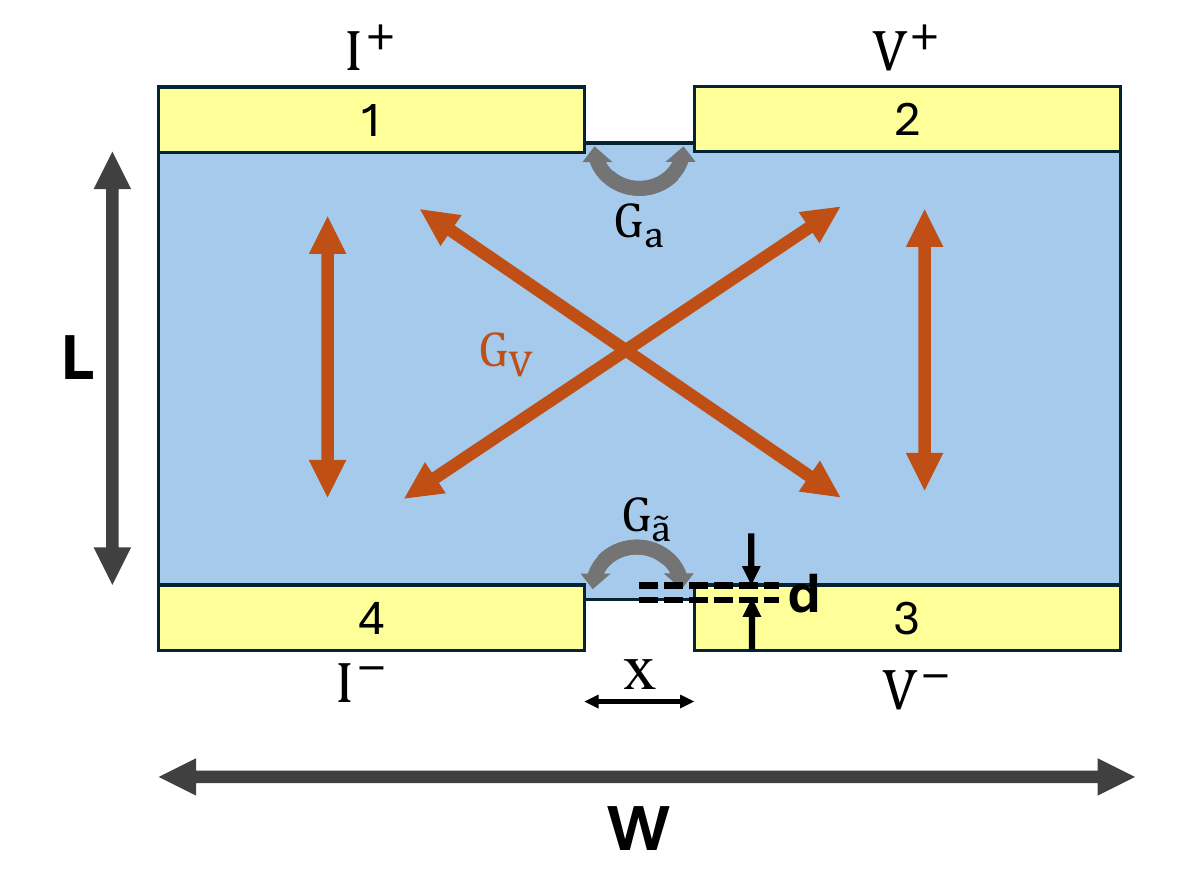}
    \caption[Multi-probe device schematic]{\textbf{Multi-probe device schematic:} The hBN-encapsulated graphene channel (blue) is connected to 4 electrical contacts labeled with $1,~2,~3,$ and $4$ (yellow). Regions $\mathfrak{R}_\mathrm{I+}$, $\mathfrak{R}_\mathrm{I-}$, $\mathfrak{R}_\mathrm{V+}$, $\mathfrak{R}_\mathrm{V-}$ and $\mathfrak{R}_\mathrm{channel}$ are given by region 1 (yellow), region 4 (yellow), region 2 (yellow), region 3 (yellow), and the blue central region, respectively.
    $L$ and $W$ are respectively the length and width of the device channel, excluding contacts. The separation between horizontally adjacent contacts is labeled by $X$, while the contacts are etched into the channel by distance $d$. The conductances between probes $1-2$ and $3-4$ are quantified by $G_\mathrm{a}$ and $G_{\tilde{\mathrm{a}}}$ (grey) respectively. The conductances through the bulk of the sample are quantified by $G_\mathrm{V}$ (orange) where the subscript $\mathrm{V}$ denotes `viscous'. }
    \label{fig:Device-Schematic}
\end{figure}

Analysis of observed variability in the experimental four-probe resistance was performed using numerical simulations. Unless otherwise mentioned, we shall consider the rectangular four-probe geometry as illustrated in Fig. \ref{fig:Device-Schematic} throughout our simulations. To simplify the problem, we work in the intermediate density regime (away from the Dirac point) and consider low-Reynolds-number, incompressible flow at steady state. Under the aforementioned assumptions, the linearized electronic Navier-Stokes equations \cite{torre2015nonlocal,guo2017higher} are given as:
\begin{gather}
-\sigma_0\bm{\nabla}\phi(\bm{r})+D_\mathrm{\nu}^2 \nabla^2\bm{J}(\bm{r}) = \bm{J}(\bm{r})
    \label{eqn:NavierStokes}\\
\bm{\nabla}\cdot\bm{J}(\bm{r}) = 0
\label{eqn:incompressibility}
\end{gather}
Here, $\sigma_0 = ne^2\tau/m^*$ is the `Drude-like' conductivity, $\phi(\bm{r})$ is the electric potential at position $\bm{r}$, $D_\mathrm{\nu} = \sqrt{\nu \tau} \sim \sqrt{l_\mathrm{ee}l_\mathrm{mr}/4}$ is the vorticity-diffusion length or the Gurzhi length and $\bm{J}(\bm{r})$ is the linearized steady-state electric current density.
Furthermore, $n$ is number density, $e$ is electronic charge, $\tau$ is the effective momentum non-conserving scattering time of a fluid element (in our simulations, $\tau = \tau_\mathrm{mr}$), $m^* = \hbar k_\mathrm{F}/v_\mathrm{F}$ is the cyclotron mass, $\nu = \eta /m^*n$ is the kinematic viscosity of the electron fluid given by ratio of shear viscosity $\eta$ to mass density $\rho = nm^*$. Note that from Eq. \ref{eqn:NavierStokes}, in the limit $D_{\nu} \rightarrow0$, we recover Ohm's law $\bm{J}(\bm{r}) = \sigma_0 \bm{E}(\bm{r})$, where $\bm{E}(\bm{r}) = -\bm{\nabla}\phi(\bm{r})$ is the electric field. In order to understand where the boundary conditions are imposed, we divide our geometry into regions $\mathfrak{R}_\mathrm{I+}$, $\mathfrak{R}_\mathrm{I-}$, $\mathfrak{R}_\mathrm{V+}$, $\mathfrak{R}_\mathrm{V-}$ and $\mathfrak{R}_\mathrm{channel}$ (see Fig. ~\ref{fig:Device-Schematic}) representing the corresponding areas of current probes, voltage probes and the channel (does not include probes) respectively. The corresponding boundaries are therefore represented by $\bm{r}_\mathrm{I+}$, $\bm{r}_\mathrm{I-}$, $\bm{r}_\mathrm{V+}$, $\bm{r}_\mathrm{V-}$ and $\bm{r}_\mathrm{channel}$. It must be specified that $\bm{r}_\mathrm{channel}$ represents the channel boundaries that are not in common with the probe boundaries. 
\begin{itemize}
\item Boundary conditions at the current contacts:
    \begin{gather}
        \phi(\bm{r}_\mathrm{I\pm}) = \pm I_0/\sigma_0 
        \label{eqn:current_contact_potential}
        \\
        \hat{n}\cdot\bm{J}(\bm{r}_\mathrm{I\pm}) = \pm I_0/(a+d) = \pm J_0
        \label{eqn:current_contact_flux}
    \end{gather}
    Here, the contacts are rectangular with length $a = (W-X)/2$ and breadth $d$ that is partly inside the channel as shown in Fig. \ref{fig:Device-Schematic}, and $\hat{n}$ represents the unit vector that is normal to the boundary. $I_0$ is the net current injected through the current probes, and $J_0$ is the electric current density at the current probes.
    Eq. \ref{eqn:current_contact_potential} implies that the electric potential at the current contacts is fixed by Ohm's law, and Eq. \ref{eqn:current_contact_flux} implies that the net flux of the current density at the current contacts is given by the total current injected divided by the total length of the contact surfaces through which current is injected, which simply follows from the definition of current density. 
      
    \item Boundary conditions at the voltage contacts and the channel boundaries:
    \begin{itemize}
    \item \textbf{Zero flux} means that the normal component of the current density vanishes at the boundary. We use the equation:
    \begin{equation}
        \hat{n}\cdot\bm{J}(\bm{r}) = 0 ~|~\bm{r} = \bm{r}_\mathrm{channel}, \bm{r}_{V\pm}
    \end{equation}

    \item \textbf{Finite-slip length} is a general treatment of the tangential current density at the boundary.  
    Due to the rough nature of the physical device boundaries, the electron fluid experiences an opposing tangential frictional force at the device edges. An increase in the frictional force exerted by the device edges on the electron fluid reduces the tangential component of the electron fluid's velocity at the boundaries. In the linear response regime, this effect can be accommodated and quantified by the parameter $l_\mathrm{b}$, known as the boundary slip-length. The following equation is used in consideration of slip-length \cite{bandurin2016negative, torre2015nonlocal}:
    \begin{equation}
    \label{eqn:finite-lb}
        \varepsilon_{ij} n_i n_k(\partial_jJ_k(\bm{r}) + \partial_kJ_j(\bm{r}))= \varepsilon_{ij}n_i J_j/l_\mathrm{b} ~|~\bm{r} = \bm{r}_\mathrm{channel}, \bm{r}_{V\pm}
    \end{equation}
    
    Where $\varepsilon_{ij}$ is the 2D Levi-Civita tensor, $n_i$ is the $i$th component of the unit normal vector to the boundary, $J_j$ is the $j$th component of the current density, and $\partial_k$ is the partial derivative with respect to $x_k$. Note that we use the convention for coordinate indices as $(x,y) = (x_1, x_2)$, so indices take values from $\{1,2\}$. Summation over repeated indices is implied in Eq. \ref{eqn:finite-lb}.
    
    The two extreme cases of slip boundary conditions, i.e. $l_\mathrm{b}\rightarrow0$ and $l_\mathrm{b}\rightarrow\infty$ are described below.
    \subitem (a) \underline{No-slip} ($l_\mathrm{b}\rightarrow0$) boundary conditions imply that the current density goes to zero at the edges of the sample. The equation is given by:
    \begin{equation}
        {\bm{J}(\bm{r})= \bm{0}~|~\bm{r} = \bm{r}_\mathrm{channel}, \bm{r}_{V\pm}}
        \label{eqn:no_slip}
    \end{equation}
    
    \subitem (b) \underline{Perfect-slip} ($l_\mathrm{b}\rightarrow\infty$, used for specific cases and replaces no-slip boundary conditions when imposed) boundary conditions imply that there is zero tangential shear stress at the boundary. The equation is:
    \begin{equation}
        \partial_yJ_x(\bm{r}) + \partial_xJ_y(\bm{r})= \bm{0}~|~\bm{r} = \bm{r}_\mathrm{channel}, \bm{r}_{V\pm}
        \label{eqn:perfect_slip}
    \end{equation}

    \end{itemize}
    Note that the zero flux condition at the voltage contacts and channel boundary deals with the normal current density at the boundaries, while the slip length conditions deal with tangential current density at the boundaries. As such, they are independent conditions. In our simulations, zero flux is always enforced at the voltage contacts and channel boundaries, irrespective of the value of $l_\mathrm{b}$.
    
\end{itemize}
Numerical simulations based on the finite element method (FEM) were performed with the above equations and boundary conditions. The results are described in subsequent sections. 

\subsection{Dependence of current flow profile in 4-probe architecture on device dimensions}
\begin{figure}[tbh]
    \centering
    \includegraphics[width=1\linewidth]{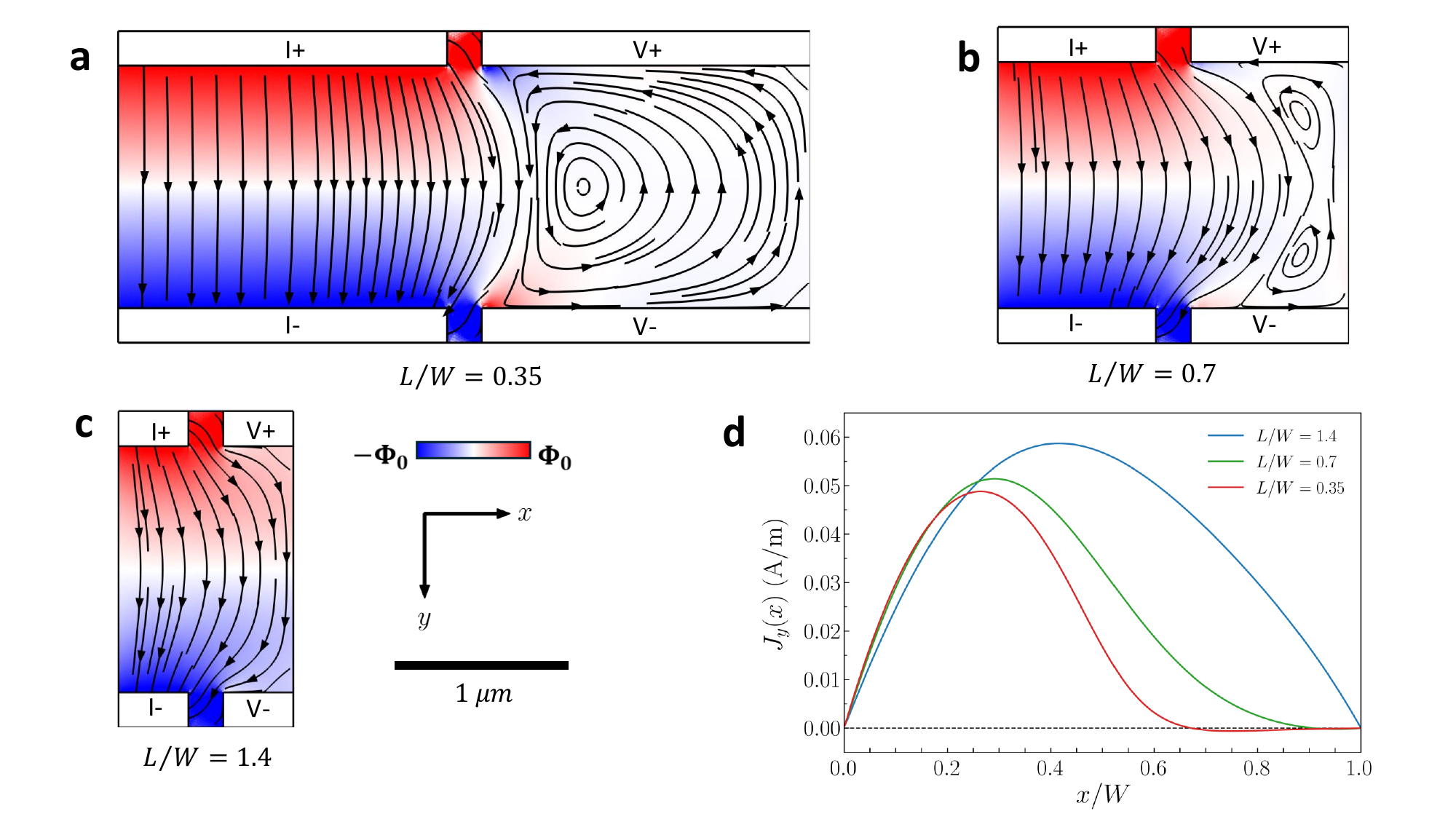} 
   \caption[Numerical simulations of 4-probe device architecture with varying $L/W$ ratios]{\textbf{\boldmath Numerical simulations of 4-probe device architecture with varying $L/W$ ratios:} (a) 4-probe device with $L = 1.4 \ \mathrm{\mu m}$ and $W = 4 \ \mathrm{\mu m}$. The red-blue surface plot represents electric potential $\phi(\bm{r})$ with black streamlines of current density $\bm{J}(\bm{r})$. The direction of $\bm{J}(\bm{r})$ is indicated by black arrows on streamlines. $\boldsymbol{\Phi_0} = I_0/\sigma_0$ in the legend. A large vortex is seen between the voltage probes on the right-hand side of the device. (b) Same plot as (a) with $L = 1.4 \ \mathrm{\mu m}$ and $W = 2 \ \mathrm{\mu m}$. Two small vortices start to appear between the voltage probes. (c) Same as (a) with $L = 1.4 \ \mathrm{\mu m}$ and $W = 1 \ \mathrm{\mu m}$. Here, the width of the device is too small for vortices to develop. (d) Flow profile $J_y(x)$ as a function of $x/W$ obtained from the simulations in (a)---(c), at constant $y$-coordinate $L/2$. Note that the positive $y$-axis is taken to be in the downward direction. Therefore, the origin of our coordinate system is at the top left corner of the rectangular device.
    }
    \label{fig:LbyW_4probe}
\end{figure}

Numerical simulations of our 4-probe device architecture were performed with varying $L/W$ ratios (see Fig. \ref{fig:LbyW_4probe}). In these simulations, $L$ was fixed at $1.4 \ \mathrm{\mu m}$, while $W$ took values $1 \ \mathrm{\mu m}$, $2 \ \mathrm{\mu m}$, and $4 \ \mathrm{\mu m}$. In Fig. \ref{fig:LbyW_4probe}a and Fig. \ref{fig:LbyW_4probe}b, we see vortex formation in the channel between the voltage probes. However, in Fig. \ref{fig:LbyW_4probe}c we do not see a vortex. Because of the small width, there is not enough space for the shear flow to generate a back-circulating vortex. From Fig. \ref{fig:LbyW_4probe}d, we observe that the vertical current flow profile $J_y$ as a function of $x$ is similar to Poiseuille flow for $x/W < 0.5$. This is in the region to the left of the device, between the current probes, where the vortex is absent. Therefore, in this region, we would expect an approximately parabolic profile as the only features of the flow are shear stress between adjacent layers with a nearly constant electric field. However, in the region $x/W > 0.5$, the vortex induces backflow, and hence the current profile deviates sharply from Poiseuille flow. We see in Fig. \ref{fig:LbyW_4probe}d from the red curve ($L/W = 0.35$) that $J_y$ becomes negative around $x/W = 0.75$, which we expect due to the formation of a vortex in that region. In general, we can expect Poiseuille-like flow in the region directly in between the current probes ($x<$~contact width), but for larger $x$ we will start to see vortex formation and a deviation from Poiseuille flow.

\subsection{Dependence of Poiseuille flow on device dimensions } \label{Poiseuille L/W}
\begin{figure}[tbh]
    \centering
    \includegraphics[width=1\linewidth]{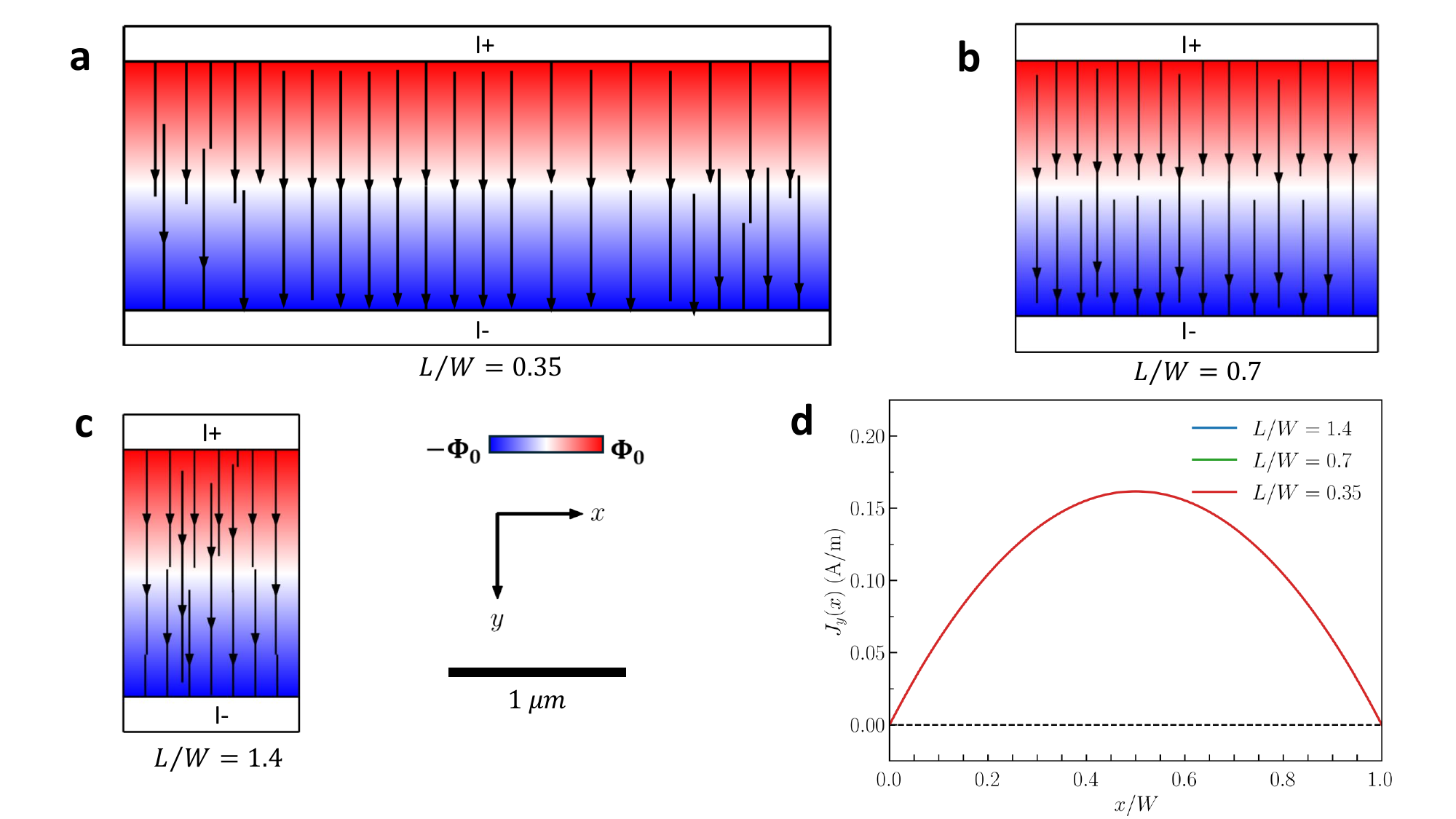} 
    \caption[Dependence of Poiseuille flow on device dimensions]{\textbf{Dependence of Poiseuille flow on device dimensions:} (a) Numerical simulations of the device with $L = 1.4$~$\mu$m and $W = 4$~$\mu$m. The red-blue surface plot represents electric potential $\phi(\bm{r})$ with black streamlines of current density $\bm{J}(\bm{r})$. The direction of $\bm{J}(\bm{r})$ is indicated by black arrows on streamlines. $\boldsymbol{\Phi_0} = I_0/\sigma_0$ in the legend.  (b) Same plot as (a) with $L = 1.4$~$\mu$m and $W = 2$~$\mu$m. (c) Same plot as (a) with $L = 1.4$~$\mu$m and $W = 1$~$\mu$m. (d) Flow profile as a function of $x$-coordinate along the width of the device for the various $L/W$ ratios in (a)---(c). We observe that all 3 curves coincide exactly, as discussed in Section \ref{Poiseuille L/W}. Note that the positive $y$-axis is taken to be in the downward direction. Therefore, the origin of our coordinate system is located at the top-left corner of the rectangular device.
    \label{fig:L_by_W}}
\end{figure}

In our simulations (refer to Fig. \ref{fig:L_by_W}), Poiseuille flow is observed for a wide range of rectangular geometries or $L/W$ ratios. The only relevant parameter in this regard is the hydrodynamic entrance length $L_h$ – the distance the electrons must travel through the channel after being injected through the contacts to achieve 99\% of the velocity flow profile \cite{fluid-mechanics-hydrodynamic-entrance-length}. Using the expression derived in previous works \cite{hydroynamic-entrance-length-rectangle}, this length can be estimated using the expression $L_h=\ KD_eRe$ where $K$ is a constant depending on the geometry of the channel, $D_e$ is the diameter/width of the channel and $Re$ is the Reynolds number of the flow. Using the expression for Reynolds number for a 4-probe architecture as given in Ref. \cite{torre2015nonlocal}, we get $Re = I_0/en\nu$. Using typical values for these parameters in graphene samples, $I_0 = 100\ \mathrm{nA}, \ n = 10^{11}~\mathrm{cm^{-2}}, \ \nu = 10^3\ \mathrm{cm^2 ~s}$ \cite{bulkandshear}, yields a Reynolds number $Re \sim 0.01 \ll 1$. This justifies our linearization of the Navier-Stokes equations. Using $K = 0.05$ \cite{Bergman2011}, $D_e = 5 \ \mathrm{\mu m}$, and the above calculated value of $Re$ in our device to calculate $L_h$, we get $L_h = 0.05 \times 5 \times 0.01\,\mu\mathrm{m} = 2.5\,\mathrm{nm} \ll W$ for all values of $W$ in our devices. This implies that the flow takes a negligible distance to develop. Hence, we can assume the flow to be fully developed at all points in the channel.

 Although the calculations of $L_\mathrm{h}$ suggest that the length and width of our samples have the required dimensions to establish a Poiseuille flow profile, we note that the above-mentioned equation ($L_h=\ KD_eRe$) was derived for a liquid of molecules/atoms flowing through a macroscopic pipe.
In addition to this, the theoretical results derived in Ref. \cite{torre2015nonlocal} were calculated in the $W/L\rightarrow\infty$, i.e. $L/W\rightarrow0$ limit, and electrical transport experiments \cite{bandurin2016negative} probing Dirac fluids have also been performed in devices with a contact geometry enabling $L/W\ =\ 0.4$. We have also performed numerical simulations of electric potential and current distributions to understand the electron flow profile in our device architecture, as shown in Fig. \ref{fig:L_by_W} for three different values of $L/W$. The corresponding current flow profiles are also presented in Fig. \ref{fig:L_by_W}, and they are all Poiseuille-like, irrespective of the aspect ratio. This is expected, as once again, the only relevant parameter for the development of the flow is the hydrodynamic entrance length $L_h \sim 2.5 \ \mathrm{nm}$, which is far less than the device dimensions. Because of this, we expect the parabolic flow to be fully developed throughout the sample. Hence, the vertical current density profile $J_y(x)$ plotted against the $x$-coordinate normalised by the width of the sample is expected to coincide for all 3 ratios of $L/W$, as seen in Fig. \ref{fig:L_by_W}d. 

\subsection{Dependence of 4-probe resistance on distance between contacts}

\begin{figure}[tbh]
    \centering
    \includegraphics[width=1\linewidth]{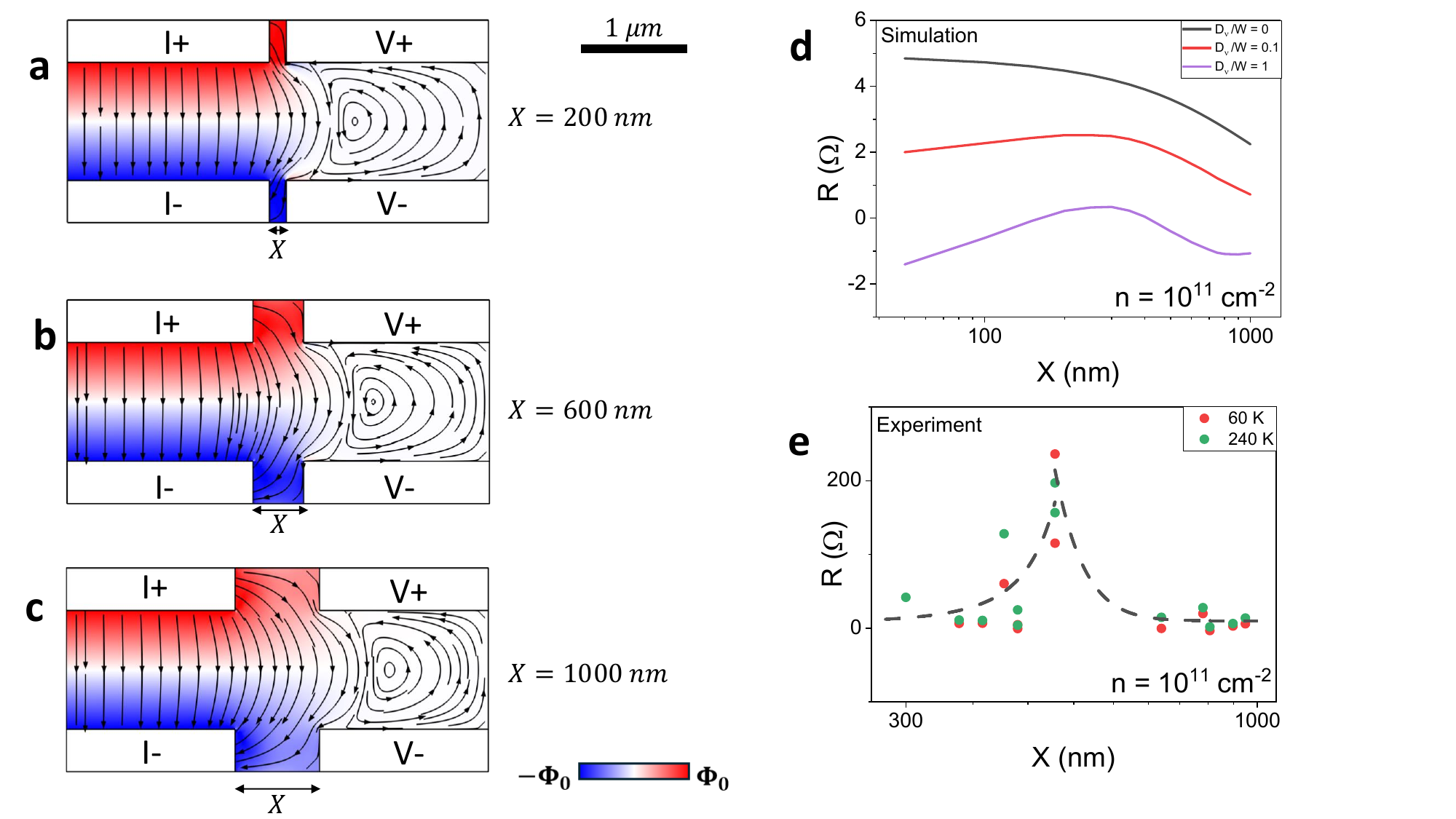} 
    \caption[Dependence of $R_{4p}$ on $X$]{\textbf{Dependence of $R_{4p}$ on $X$:} (a) --- (c) Numerical simulations of our device with varying $X$. Device dimensions are $L = 1.4 \ \mathrm{\mu m}$ and $W = 5 \ \mathrm{\mu m}$. $D_{\nu}$ was set to $W/10 = 0.5 \ \mathrm{\mu m}$. (d) Plot of $R_{4p}$ vs $X$ obtained from simulation. Note the initial increase, followed by a decrease in the viscous regime ($D_{\nu} > 0$), compared to the monotone decreasing curve in the diffusive regime ($D_{\nu} = 0$). (e) Experimental data for variation of $R_{4p}$ with $X$ at 60 K and 240 K. The dashed black curve serves as a guide to the eye.}
    \label{fig:R_vs_x}
\end{figure}

Numerical simulations of the electric potential and current distributions in our device architecture with varying distance between adjacent contacts $X$ are presented in Fig. \ref{fig:R_vs_x}. In addition, we have compared numerical simulations of electrical resistance within our device geometry to the experimental data at $n={10}^{11} \ \mathrm{cm^{-2}}$ to illustrate the influence of $X$ on electrical transport, as shown in Fig. \ref{fig:R_vs_x}d and Fig. \ref{fig:R_vs_x}e in log-linear scale.    

We see in Fig. \ref{fig:R_vs_x}d that the exact dependence for the viscous case $(D_{\nu}/W>0)$ is non-trivial for small $X$ but continues to decrease for larger $X$. In Fig. \ref{fig:R_vs_x}e, the data points are obtained from different devices with different $L$, $W$, and $X$, and therefore do not constitute a fair comparison. However, the trend remains similar with an increase in resistance at intermediate $X$ and a decrease for high $X$. Note that the dashed curve in Fig. \ref{fig:R_vs_x}e is not a fitted curve, but a guide to the eye illustrating the trend followed. 

\section{Effect of slip length and Gurzhi length on hydrodynamic transport}
The most general treatment of finite slip length boundary conditions is defined by Eq. \ref{eqn:finite-lb}. However, as argued in section S5, it is natural to use no-slip or low-slip ($l_\mathrm{b} \ll W$) boundary conditions. An analytical expression for longitudinal conductivity $\sigma_{xx}$ for a straight channel with slip length $l_\mathrm{b}$ has been derived in Ref.~\cite{torre2015nonlocal}. For a channel of width $W$, ohmic conductivity $\sigma_0$, and Gurzhi length $D_\mathrm{\nu}$, the longitudinal conductivity is
\begin{equation}
\sigma_{xx} = \sigma_0 \left[1 -\frac{2 D_\mathrm{\nu}^{2} \sinh\!\left( \frac{W}{2D_\mathrm{\nu}} \right)}{W\!\left\{ D_\mathrm{\nu} \cosh\!\left( \frac{W}{2D_\mathrm{\nu}} \right) + l_\mathrm{b} \sinh\!\left( \frac{W}{2D_\mathrm{\nu}} \right) \right\}}\right]
\end{equation}

Under the approximations $D_\mathrm{\nu} > W/2$ and $D_\mathrm{\nu} > l_\mathrm{b}$, and keeping only the first-order correction, the following equation is obtained:
\begin{equation}
\sigma_{xx}
=
\frac{n^{2} e^{2} W^{2}}{12\,\eta}
\left(
1 + \frac{6l_\mathrm{b}}{W}
\right)
\end{equation}
where $\eta$ is the dynamic viscosity of the electron fluid, $e$ is the electronic charge, and $n$ is the carrier density.

\begin{figure}[tbh]
    \centering
    \includegraphics[width=1\linewidth]{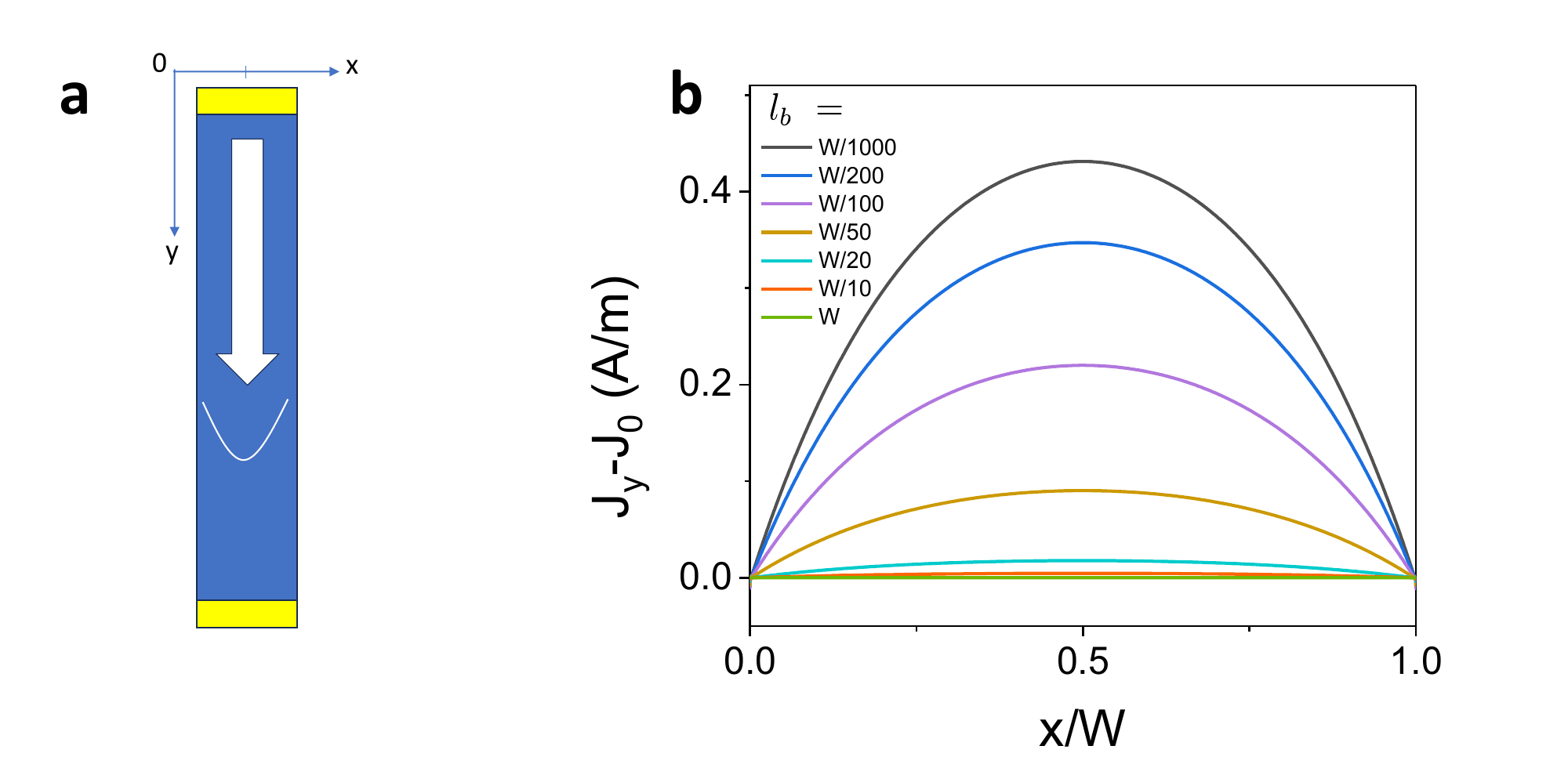} 
    \caption[Hydrodynamic-ballistic crossover with variation of slip length]{\textbf{Hydrodynamic-ballistic crossover with variation of $\bm{l_\mathrm{b}}$} (a) Schematic figure displaying the geometry used for the simulations in (b). (b) Plot of the $y$-component of the current density $J_y$ as a function of $x/W$ in a straight channel geometry of width $W=5 \ \mathrm{\mu m}$ for varying slip length $l_\mathrm{b}$. The plots were obtained from numerical simulations with the geometry described in the schematic figure (a). Here, $J_0$ is the current density at $x/W=0$. 
    \label{fig:Lb-variation}}
\end{figure}

Thus, the correction introduced by a nonzero slip length $l_\mathrm{b} \ll W$ is very small and may be neglected.
The variation of the current flow profile in a straight channel geometry of width $W = 5\,\mu\mathrm{m}$ (shown in the schematic Fig. \ref{fig:Lb-variation}a)  as a function of slip length is provided in Fig. \ref{fig:Lb-variation}b, which clearly suggests that the sample can sustain Poiseuille-like flow for slip lengths up to $l_\mathrm{b}=W/50 = 100 ~ \mathrm{nm} $. From Fig. \ref{fig:Lb-variation}b, we see that for $l_\mathrm{b}/W >0.02$, the flow profile becomes flat, which can be attributed to the device entering the ballistic regime. Hence, for devices having a width of $\sim 5~\mu$m, viscous electronic transport can persist in the channel, up to a slip length of $l_\mathrm{b}<100 \ \mathrm{nm}$.

\begin{figure}[tbh]
    \centering
    \includegraphics[width=1\linewidth]{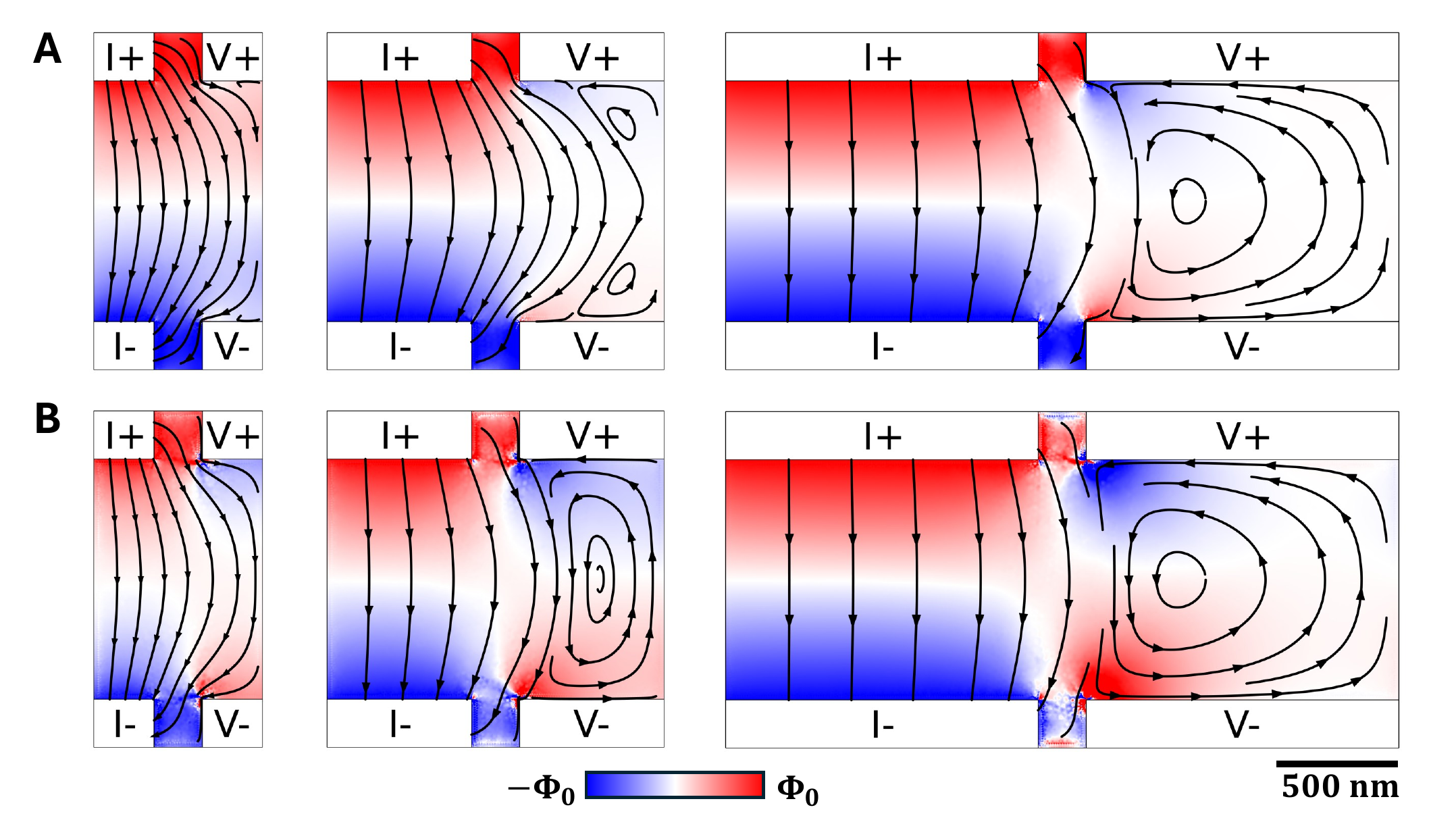}
    \caption[Comparison of no-slip and perfect slip boundary conditions for different $L/W$ ratios]{\textbf{Comparison of no-slip and perfect slip boundary conditions for different $L/W$ ratios}: Red-blue surface plot of electric potential $\phi(\bm{r})$ with black streamlines of current density $\bm{J}(\bm{r})$ with no-slip (A) and perfect-slip (B) boundary conditions. The direction of $\bm{J}(\bm{r})$ is indicated by black arrows on streamlines. Different columns (from left to right) in (A) and (B) represent $L/W = 1.4,~0.7$ and $0.35$, respectively. In these simulations, $X = d = 200 \ \mathrm{nm}$ and $L$ is $1 \ \mathrm{\mu m}$. $\boldsymbol{\Phi_0} = I_0/\sigma_0$ in the legend and the scalebar is $500\ \mathrm{nm}$.
    }
    \label{fig:simulation2}
\end{figure}

Furthermore, to understand the role of boundary conditions and the aspect ratio on the hydrodynamic flow (i.e. when $D_\mathrm{\nu} \ne 0 \ \mathrm{\mu m}$), different aspect ratios $L/W>0.5$ and $L/W<0.5$ with both limiting cases of $l_b \rightarrow 0 $ (no-slip) and $l_b \rightarrow \infty $ (perfect slip) have been plotted in Fig. \ref{fig:simulation2}. We observe that the perfect slip boundary condition yields a stronger voltage reversal on the right side of the device compared to the no-slip condition. However, the qualitative features are largely independent of the choice of boundary conditions, except for the figures corresponding to $L/W = 1.4$. Here, the voltage reversal is observed for the perfect slip boundary conditions, whereas this is not the case for the no-slip boundary conditions. In simulations, no-slip boundary conditions exhibit stronger Poiseuille-like behaviour than free slip boundary conditions (see Fig. \ref{fig:Lb-variation}), which is expected as no-slip or low slip (small $l_\mathrm{b}$) is required for Poiseuille flow to develop in a channel.

\subsection{Dependence of electric current and potential profile on Gurzhi length}
\begin{figure}[tbh]
    \centering
    \includegraphics[width=1\linewidth]{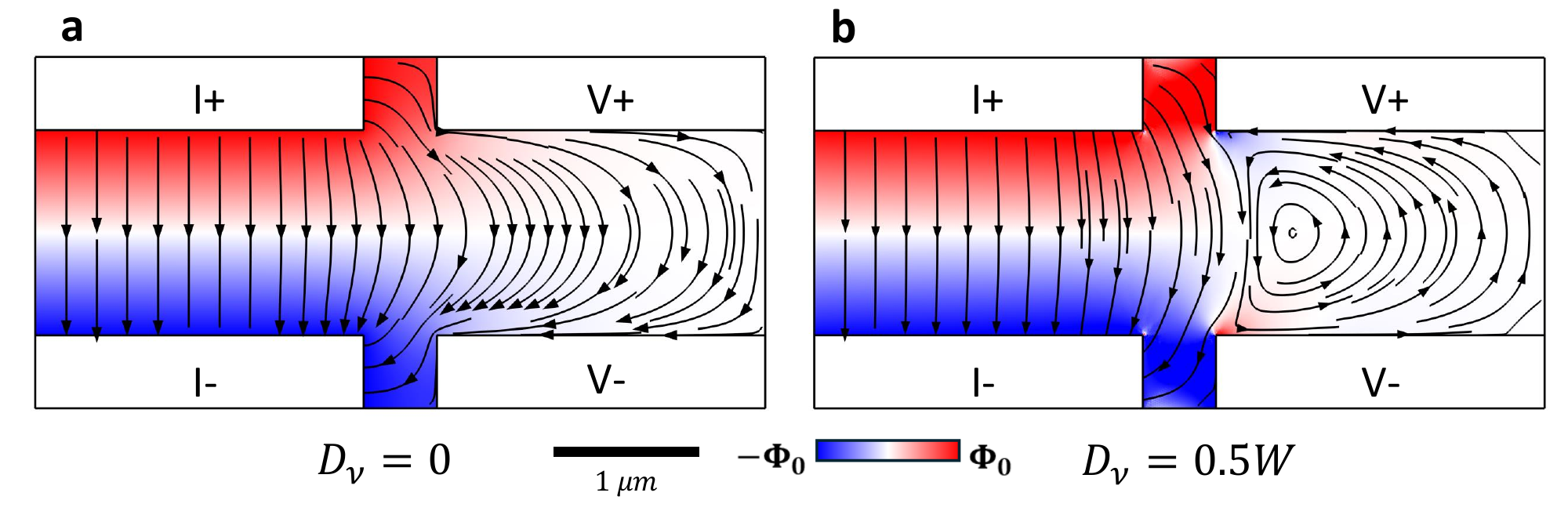} 
    \caption[Comparison of electric potential and current profiles between the diffusive and hydrodynamic regimes]{\textbf{Comparison of electric potential and current profiles between the diffusive and hydrodynamic regimes:} (a) Numerical simulations of a 4-probe geometry device in the diffusive regime ($D_{\nu} = 0$). Device dimensions are $L = 1.4 \ \mathrm{\mu m}$ and $W = 5 \ \mathrm{\mu m}$. The red-blue surface plot represents electric potential $\phi(\bm{r})$ with black streamlines of current density $\bm{J}(\bm{r})$. The direction of $\bm{J}(\bm{r})$ is indicated by black arrows on streamlines. $\boldsymbol{\Phi_0} = I_0/\sigma_0$ in the legend. (b) Same plot as (a), but in the hydrodynamic regime ($D_{\nu} = 0.5W$). Note the presence of a vortex and the reversal in electric potential between the voltage probes.
    }
    \label{fig:electrostatic_BC}
\end{figure}

\begin{figure}[tbh]
    \centering
    \includegraphics[width=1\linewidth]{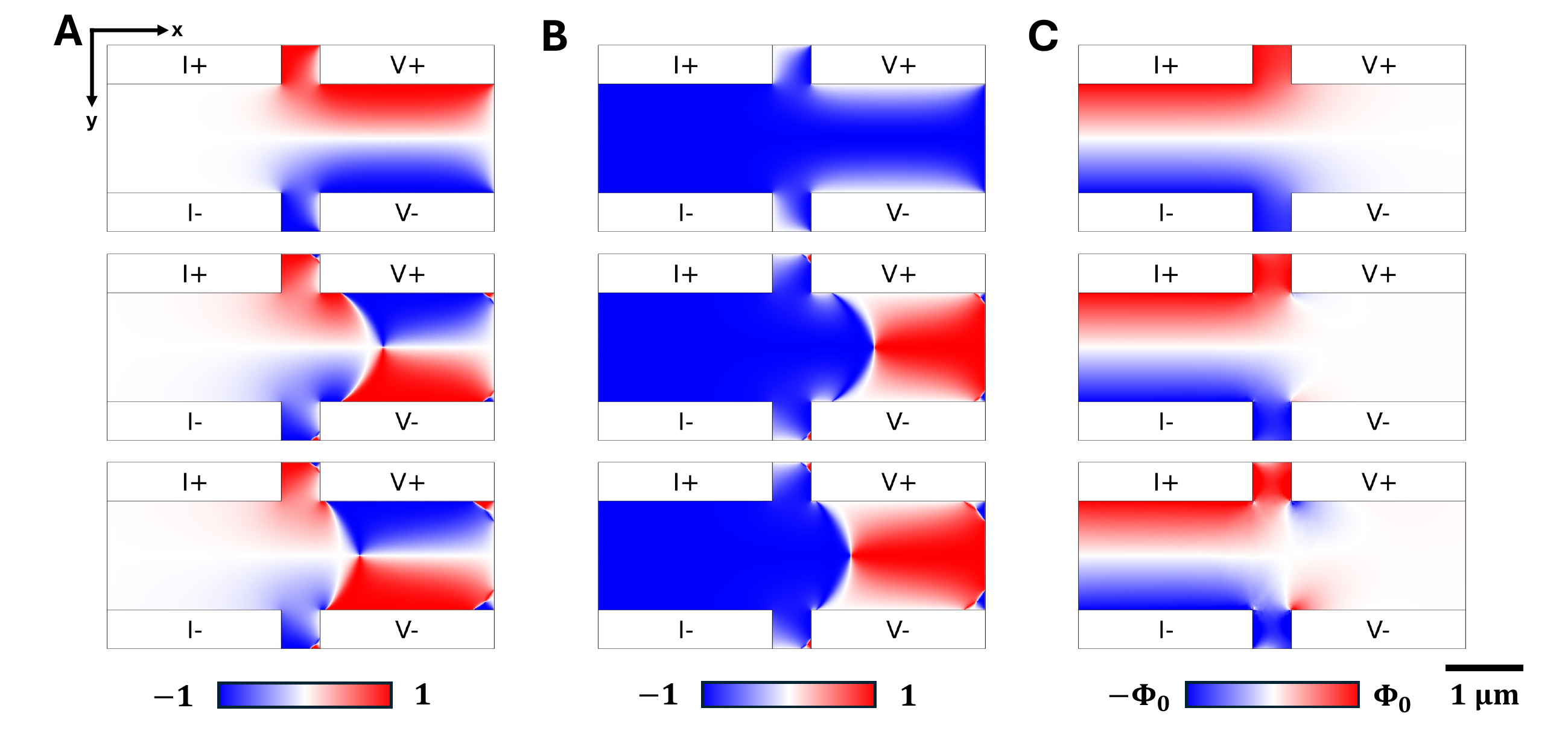}
    \caption[Comparison of $J_x,~J_y$ and $\phi$ in four-probe rectangular geometry]{\textbf{Comparison of $J_x,~J_y$ and $\phi$ in four-probe rectangular geometry}: Surface plot of normalized components of current density $J_x(\bm{r})/|\bm{J}(\bm{r})|$ in (A) and $J_y(\bm{r})/|\bm{J}(\bm{r})|$ in (B). (C) Surface plot of electric potential $\phi(\bm{r})$. $\boldsymbol{\Phi_0} = I_0/\sigma_0$ in the legend for (C), here $I_0$ is the injected current and $\sigma_0$ is the ohmic conductivity. Top to bottom: The three rows of (A), (B) and (C) represent plots for varying Gurzhi lengths (in $\mu \mathrm{m})$ $D_\mathrm{\nu} = 0,~W/10$ and $W$ respectively. In these simulations, $X = d =500 \ \mathrm{nm}$ and $L$ and $W$ are $1.4 \ \mathrm{\mu m}$ and $5 \ \mathrm{\mu m}$, respectively. Note that the sign reversal (shown in red) in $J_y$ for $D_{\nu} > 0$ between the voltage probes implies formation of a vortex, which is also seen by the reversal in potential in the region between the voltage contacts.
    }
    \label{fig:simulation1}
\end{figure}

To examine the effect of varying Gurzhi length on the potential and current flow profiles, we have performed numerical simulations of our device. We have compared simulations (see Fig. \ref{fig:electrostatic_BC}) of a fully diffusive transport regime to a hydrodynamic transport regime, as measured by the Gurzhi length $D_{\nu}$. In the diffusive ($D_{\nu}=0$) regime (see Fig. \ref{fig:electrostatic_BC}a), we do not see any vortices forming in the flow profile, but in the hydrodynamic ($D_{\nu}=0.5W$) regime (see Fig. \ref{fig:electrostatic_BC}b), we see vortex formation with a reversal in the sign of the electric potential near the voltage probes. Hence, negative resistance in this geometry, which we have observed in some of our devices, cannot be obtained as a result of diffusive transport. We have also plotted electric potential $\phi(\bm{r})$ and normalised components of the current density in Fig. \ref{fig:simulation1} with varying Gurzhi length $D_\mathrm{\nu}$. It is evident from the different plots that increase in $D_\mathrm{\nu}$ is accompanied by a sign reversal in potential in areas near voltage probes, hinting at a possible origin of the observed negative resistance.

\section{Johnson noise thermometry}

\begin{figure}[tbh]
    \centering
    \includegraphics[width=1\linewidth]{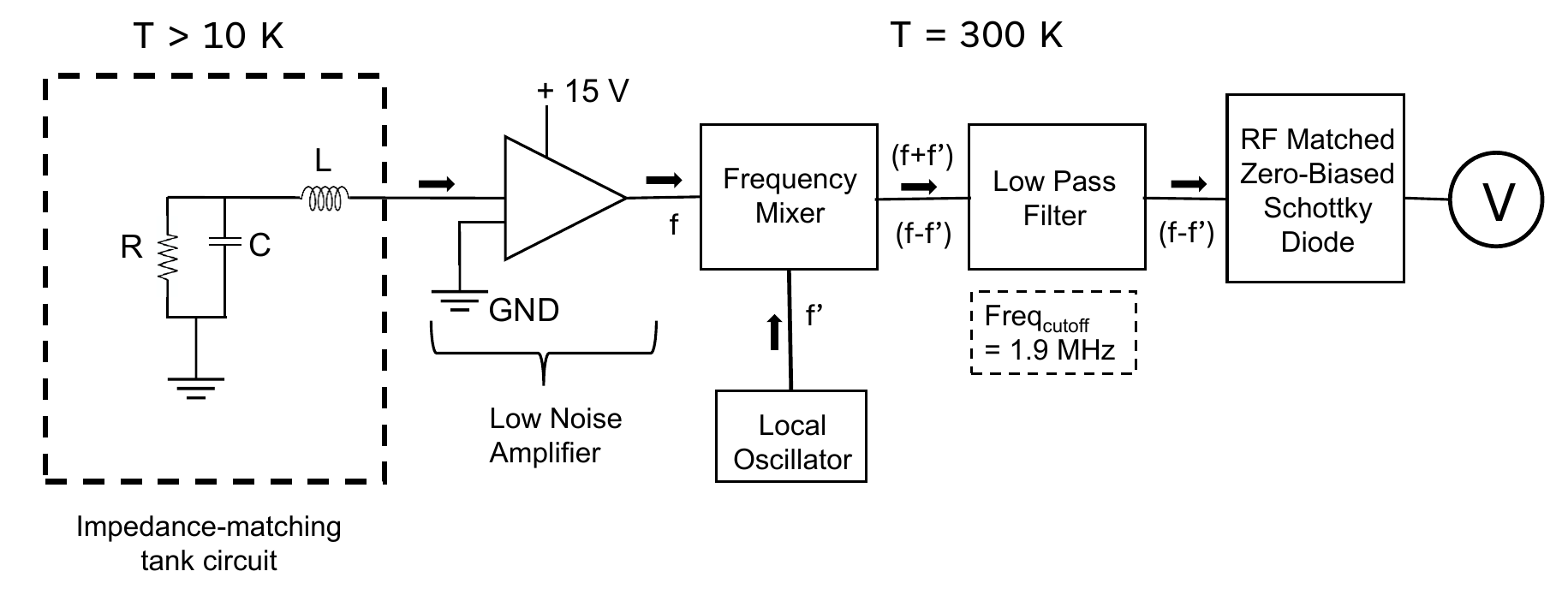}
    \caption[Johnson-Nyquist noise measurement setup]{\textbf{Johnson-Nyquist noise measurement setup:} The noise measurement circuitry consisting of cryogenic as well as room temperature components. The resistor $R$ in the circuit represents the sample.}
    \label{fig:noisecircuit}
\end{figure}

The thermal noise measurements were performed using a custom-built high-frequency noise measurement setup, as shown in Fig.~\ref{fig:noisecircuit}, where the graphene device, referred to as a resistor $R$, was wire-bonded to a printed circuit board incorporating an impedance-matched LC tank circuit.

The RMS voltage associated with the thermal noise produced by the sample can be written as
\begin{equation}
\langle V^2 \rangle = 4k_\mathrm{B} TR \Delta f 
\end{equation}

where $R$ is the two-probe resistance of the graphene device and $\Delta f$ is the measurement bandwidth.

The LC circuit, consisting of an air-core inductor $L = 220$~nH and a ceramic parallel-plate capacitor $C =3.9$~pF, when connected to the resistor $R$ yields a resonant frequency given by
\begin{equation}
f_\mathrm{0}=\frac{1}{2\pi}{\sqrt{\frac{1}{LC}-\frac{1}{R^2C^2}}}
\end{equation}

For our circuit, $f_\mathrm{0} \simeq 100$~MHz. This configuration maximises signal transmission across a narrow frequency band around $f = f_\mathrm{0}$ by ensuring optimal impedance matching at $Z = 50$~$\Omega$ and by suppressing other frequency components. The output noise power, after passing the impedance-matching $LC$ circuit, is given by

\begin{equation}
\langle V^2 \rangle = 4k_\mathrm{B} TR \Delta f \times (1-\Gamma^2)
\end{equation}

where $\Gamma$ is the reflection coefficient, given by

\begin{equation}
\Gamma = \frac{Z(\omega) - Z_0}{Z(\omega) + Z_0}
\end{equation}

with $Z(\omega)$ being the device impedance and $Z_0 = 50~\Omega$ the characteristic impedance of the transmission line. 

The impedance-matched RF signal is carried out of the cryostat through RF-compatible coaxial cables to a set of low-noise amplifiers (ZFL-500LN+ and MITEQ-AU 1291), which amplify the signal without introducing excessive thermal noise. This is quantified by the amplifiers' noise temperature $T_\mathrm{N}$. The amplified signal is subsequently transmitted into an RF mixer (ZLW-3+) driven by a local oscillator (Keysight N5173B signal generator) with a reference output signal of frequency $f_\mathrm{0}$. This mixer-oscillator combination acts as an analog down-converter, branching the signal into two frequency components: $f+f_\mathrm{0}$ and $f-f_\mathrm{0}$. A low-pass filter (SLP-1.9+) is used to transmit the downconverted component of the signal while suppressing the higher-frequency component. The resulting RMS noise voltage is given by

\begin{equation}
\langle V^2 \rangle = 4k_\mathrm{B}GS\Delta f (1 - \Gamma^2)RT
\end{equation}

where $G$ is the amplifier gain and $S$ accounts for the signal loss from the mixer and filter. The filtered signal is then passed through an RF-matched Schottky diode (DMS-104P), which integrates the signal over the low-pass filter bandwidth and converts the RF power into a DC voltage. This DC voltage, $V_\mathrm{DC}$, is finally recorded using a digital multimeter (DMM). Using information obtained from the diode calibration, the expression for the obtained DC voltage can be written as

\begin{equation}
V_\mathrm{DC}=\zeta\frac{\langle V^2 \rangle}{R}
\end{equation}

where $\zeta$ is the power-to-voltage conversion factor. This leads to

\begin{equation}
V_\mathrm{DC} = 4k_\mathrm{B} \zeta G S \Delta f (1 - \Gamma^2) \dfrac{\sum\limits_{i} T_i R_i}{\sum\limits_{i} R_i}
\end{equation}

since we can write $R = R_\mathrm{ch} + 2R_\mathrm{c}$ as the total two-probe resistance where $R_\mathrm{ch}$ is the graphene channel resistance with $T_\mathrm{ch}$ as the electronic temperature of the graphene channel and $2R_\mathrm{c}$ as the contact resistance with $T_\mathrm{c}$ is the temperature of the gold electrical contacts. This finally leads to

\begin{equation} \label{lastjneqn}
V_\mathrm{DC} = 4\alpha  G S k_\mathrm{B} \Delta f \left(1 - \Gamma^2 \right) \frac{R_\mathrm{ch}T_\mathrm{ch} + 2R_\mathrm{c}T_\mathrm{c}}{R_\mathrm{ch} + 2R_\mathrm{c}}
\end{equation}

Using Eq.~(\ref{lastjneqn}), we get the value of the electronic temperature $T_\mathrm{ch}$ of the graphene channel.

Further, we adopt the cold phonon bath approximation, wherein Joule heating of the graphene channel under DC bias is assumed to be ineffective in changing the temperature of the contacts, owing to the large thermal boundary resistance at the metal-graphene interfaces \cite{roukes1985hot} and high thermal conductivity of gold - implying that the electrical contacts stay at room temperature ($T_\mathrm{c}=296$~K). Other factors in the expressions, such as $\zeta$, $G$, $S$, $\Delta f$, are determined from the calibration procedure and known experimental settings.

The sensitivity and resolution of this technique are fundamentally limited by Dicke's radiometer formula. It sets a bound on the temperature resolution ($\delta T$) of such a noise measurement and is given by:

\begin{equation}
\delta T = \frac{T_\mathrm{sample}+T_\mathrm{N}}{\sqrt{\tau \Delta f}}
\end{equation}

where $T$ is the lattice temperature of the device ($=300$~K), $T_\mathrm{N}$ is the system noise temperature (primarily from the amplifier =$75$~K ), $\Delta f$ is the measurement bandwidth ($\simeq 4$~MHz), and $\tau$ is the integration time ($=1.5$~s). This gives $\delta T = 150$~mK at room temperature and $\delta T = 5$~mK at $10$~K. To experimentally measure the error associated with our readings, we record approximately $1000$ data points at a fixed value of the number density and the DC electric field across the channel.

\begin{figure}[tbh]
    \centering
    \includegraphics[width=0.7\linewidth]{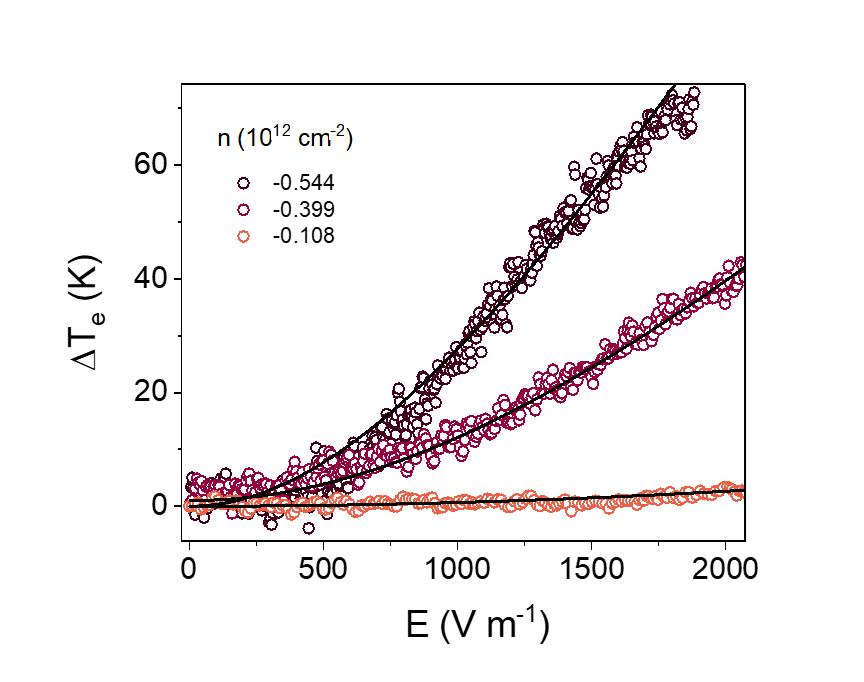}
    \caption[Bias-induced increase in electronic temperature]{\textbf{Bias-induced increase in electronic temperature:} Variation of $\Delta T_\mathrm{e}=T_\mathrm{e}(E)-T_\mathrm{e}(E=0)$ as a function of DC electric field $E$ at temperature $T=100$~K for three different carrier densities. The solid line represents the fitted curve as per Eqn.~\ref{Te}.}
    \label{fig:TeVSE}
\end{figure}

Using this circuitry, the electronic temperature of graphene has been evaluated at different temperatures and carrier densities under an in-plane DC electric field $E$. The resultant Joule heating across the electronic channel leads to a change in $T_\mathrm{e}$ as a function of $E$ (Fig.~\ref{fig:TeVSE}). Solving the heat diffusion equation \cite{majumdar2025universality} for this geometry, we get
\begin{equation}
    T_\mathrm{e} = \dfrac{2\cal L}{3L^2E^2}\left[\left(T_\mathrm{c}^2 + \dfrac{L^2}{\cal L} E^2 \right)^{3/2} - \left(T_\mathrm{c}^2\right)^{3/2}\right]
    \label{Te}
\end{equation}
where $L$, $\cal L$ and $T_\mathrm{c}$ are respectively the length of the channel, effective Lorenz number and electron temperature at the metallic contact. By fitting $T_\mathrm{e}$ against $E$ using Eq.~\ref{Te}, we estimate $\mathcal{L}$ at different densities for the electron fluid in graphene. The plot of $\mathcal{L}/\mathcal{L}_\mathrm{WF}$ as a function of $n$ for three different devices is reproducible across all our high-quality samples in pristine condition and exhibits a strong violation of the WF law. Since the charge and heat flow pathways are decoupled in the presence of strong momentum-conserving scattering, this violation is considered to be a signature of viscous electronic flow \cite{crossno2016observation, principi2015}. As per relativistic hydrodynamics, for a fluid of Dirac fermions \cite{lucas2018hydrodynamics,li2022hydrodynamic}, $\mathcal{L}/\mathcal{L}_\mathrm{WF}$ varies as 

\begin{equation}
    {\cal L}(n,T) = \frac{1}{e^2}{{\left [\frac{s(T)n_0(T)}{n^2+n^2_0(T)}\right ]}}^2
\label{L}
\end{equation}
where entropy density $s(T)$ quantifies the thermal disorder of the Dirac fluid and $n_0(T)$ refers to an effective density scale, which determines the intrinsic conductivity of the electron fluid \cite{li2022hydrodynamic, li2020hydrodynamic, majumdar2025universality}. We observe a very good agreement between the experimental data and Eq. \ref{L}, suggesting Dirac fluid-like electron flow in all three graphene FETs (Fig. 2 of the main text) near the CNP.

\section{Comparison between different electrical transport regimes}
\label{sec:comparison-transport}
\begin{figure}[tbh]
    \centering
    \includegraphics[width=1\linewidth]{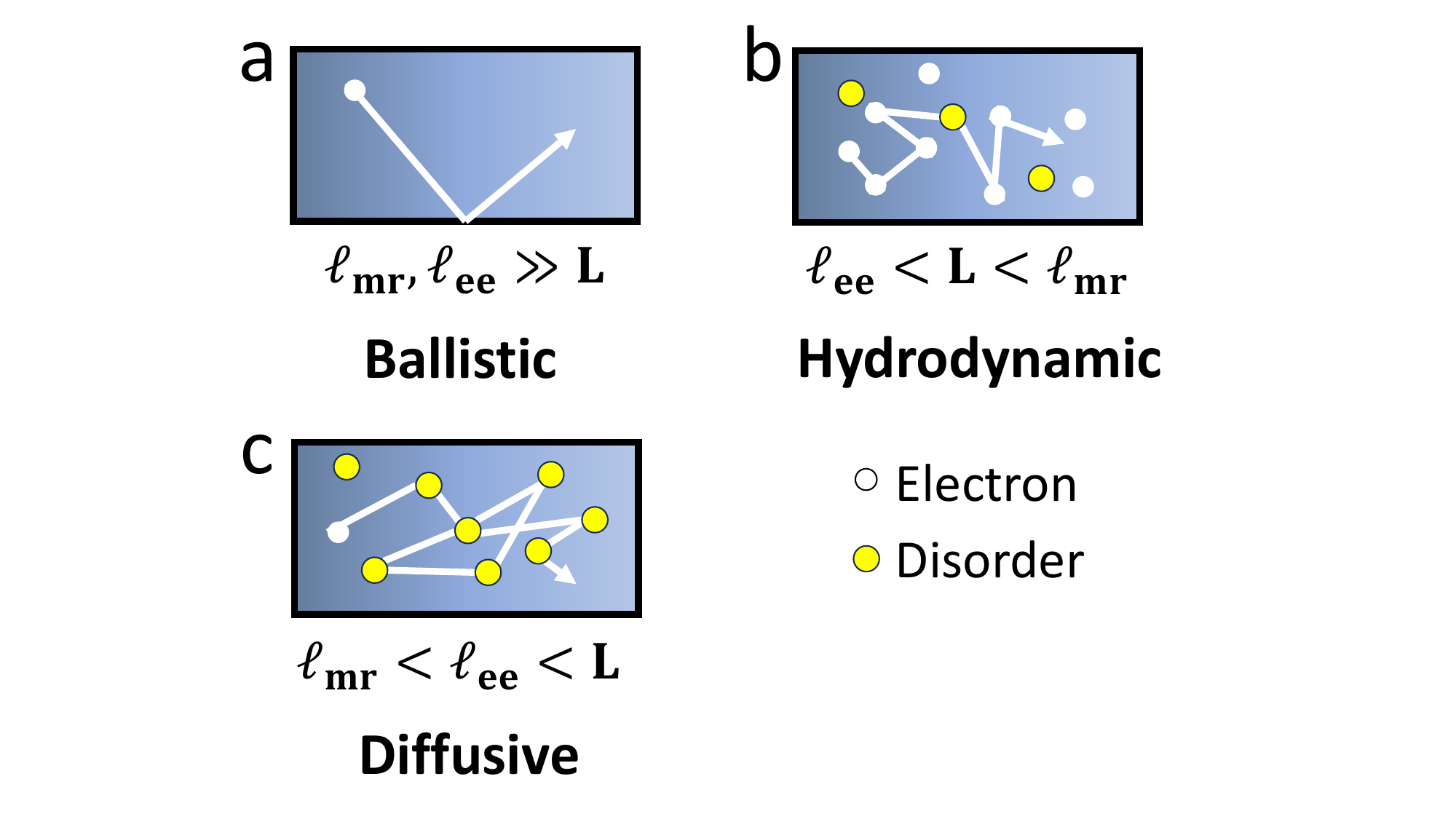}
    \caption[Schematic of different electrical transport regimes]{\textbf{Schematic of different electrical transport regimes:} (a) Schematic figure depicting ballistic transport in a rectangular channel of length $L$. Here, electrons scatter off the boundaries without undergoing collisions with other electrons or impurities. (b) Schematic figure depicting hydrodynamic transport. Here, momentum-conserving electron-electron collisions dominate, giving rise to a viscous flow of electrons. (c) Schematic figure depicting diffusive transport, where impurity and phonon scattering dominate.
    } 
    \label{fig:hydrodynamic schematic}
\end{figure}

Electrical transport in a conductor can be broadly described by comparing momentum-conserving length scales ($l_\mathrm{mc}$), where the charge carriers exchange momentum amongst themselves, and momentum-relaxing length scales ($l_\mathrm{mr}$), where the charge carriers lose their momentum to foreign entities such as defects and disorder. The shortest length-scale among the two usually governs the transport properties. In conventional metals, $l_\mathrm{mr}$ is usually the shortest length scale because electrons lose their momentum primarily by scattering due to phonons, impurities, and other disorder. However, in clean samples with low disorder, momentum-conserving collisions can dominate. In this scenario, the whole electron system's momentum remains conserved and $l_\mathrm{mc}=l_\mathrm{ee}$, where $l_\mathrm{ee}$ is the electron-electron scattering length. Such a system can be treated as a fluid, and its macroscopic quantities may be obtained using the Navier-Stokes equations. Hence, depending on the ratios of interaction length scales, electronic transport in solids can be classified into the following regimes \cite{lucas2018hydrodynamics} (see Fig. \ref{fig:hydrodynamic schematic}):

\begin{itemize}
    \item \textbf{Ballistic:}
    In the ballistic regime, the mean-free path is much larger than the characteristic length, i.e. $l_\mathrm{mfp}\gg W$ (where $l_\mathrm{mfp}$ is the mean free path and $W$ is the width or the characteristic length of the sample). At most, it results in electrons scattering off the sample boundaries; ideally, there is no scattering, and the electrons move independently. In the Fermi liquid regime, since $l_\mathrm{ee}$ diverges as $1/T^2$ as $T \rightarrow 0$ \cite{lucas2018hydrodynamics}, we expect to see ballistic transport, provided the sample has very low impurity concentration. The ballistic conductivity $\sigma$ is proportional to $(e^2/h)\cdot k_\mathrm{F}L$ \cite{guo2017higher}, where $k_\mathrm{F}$ is the Fermi wavevector, $e$ is the electronic charge, $h$ is Planck's constant, and $L$ is the length of the sample. Ballistic transport is analogous to Knudsen flow in gas dynamics, which occurs in rarefied gases. It can be quantified by the Knudsen number $\mathrm{Kn} = l_\mathrm{mfp}/W$ where $W$ is the characteristic length of the system and $l_\mathrm{mfp}$ is the mean-free path \cite{Knudsen,Bergman2011}. $\mathrm{Kn}\gg10$ indicates the free-molecular regime and is similar to the ballistic regime. Furthermore, $0.01<\mathrm{Kn}<0.1$ indicates the slip-flow regime, where the gas slips at the walls; this can be quantified by the boundary-slip length $l_b$. Increasing $l_b$ effectively leads the profile to develop plug-like flow (see Fig. \ref{fig:Lb-variation}), where the flow profile is flattened. 
    \item \textbf{Hydrodynamic:}
     In the hydrodynamic regime, for pristine samples, the momentum-conserving electron-electron collisions dominate over momentum relaxing electron-impurity or electron-phonon collisions, i.e. $l_{ee} \ll l_{mr}, W$. The viscous conductivity $\sigma$ is proportional to $W^2/\eta$, where $\eta$ is the shear viscosity. In this regime, we observe Poiseuille flow, where the flow profile ($J_y$ vs $x$) has a parabolic shape (see Fig. \ref{fig:Lb-variation}). This is the continuum regime quantified by $\mathrm{Kn} \ll 0.01$.
    \item \textbf{Diffusive:}
     In the diffusive regime, we have $l_{mr} \ll l_{ee}, W$, which results in regular ohmic behaviour. The diffusive conductivity $\sigma$ is given by the Drude relation $\sigma = (ne^2\tau/m)$, where $n$ is number density, $e$ is electronic charge and $\tau$ is the average scattering time.      
\end{itemize}

\section{Signatures of viscous flow} 
\begin{figure}[tbh]
    \centering
    \includegraphics[width=1\linewidth]{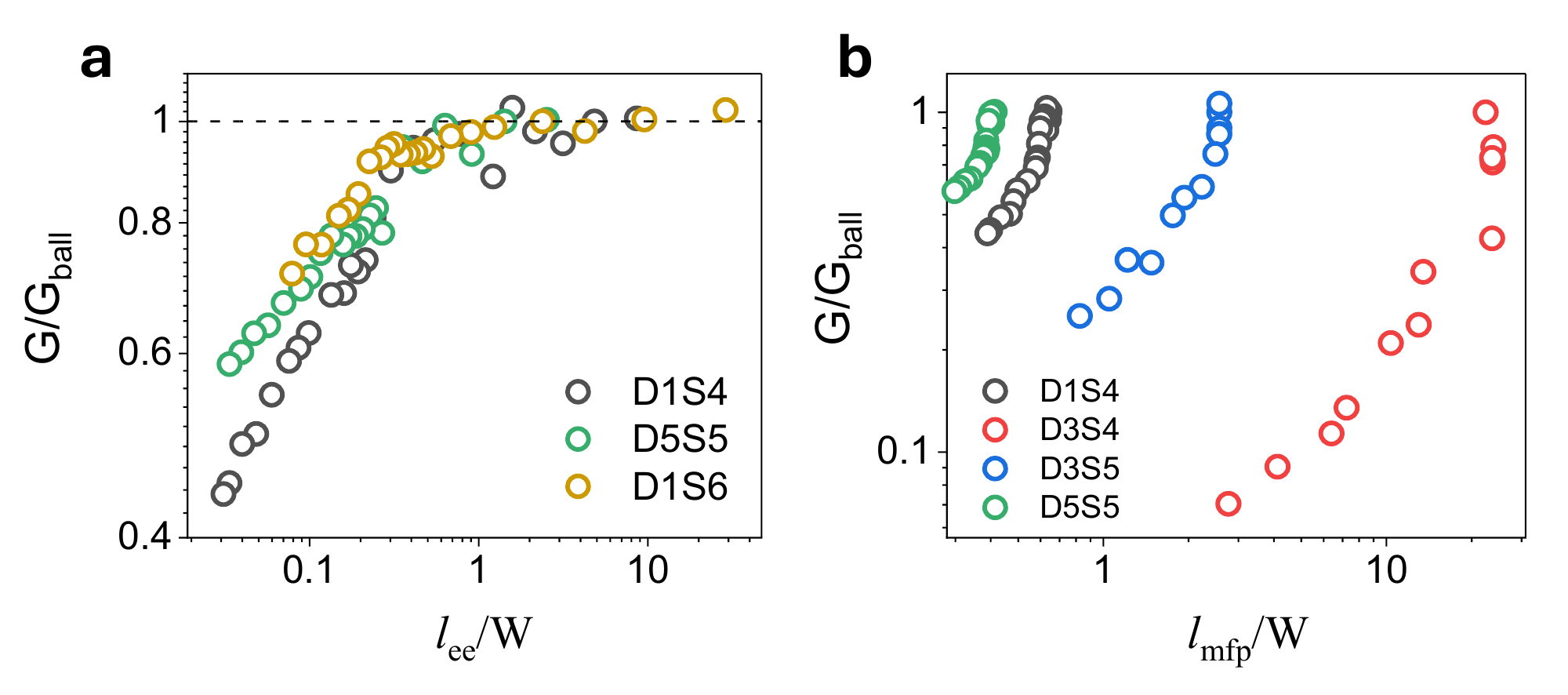}
    \caption[Hydrodynamic electron transport in graphene]{\textbf{Hydrodynamic electron transport in graphene:} (a) Electrical conductance $G$ (normalized by the ballistic conductance $G_\mathrm{ball}$) vs the ratio $l_\mathrm{ee}/W$ for devices D1S4, D5S5 and D1S6. The dashed line indicates $G/G_\mathrm{ball}=1$. The ratio $G/G_\mathrm{ball}$ saturates to 1 as we increase the ratio $l_\mathrm{ee}/W$. (b) Electrical conductance $G$ (normalized by the ballistic conductance $G_\mathrm{ball}$) vs the ratio $l_\mathrm{mfp}/W$ for devices D1S4, D3S4, D3S5, and D5S5.
    } 
    \label{fig:G-ballistic}
\end{figure}

The long electronic mean free path in our devices does not directly imply ballistic transport; rather, it indicates the suppression of impurity scattering in this regime. This, in turn, implies the dominance of either electron--electron scattering or ballistic transport. The quadratic dependence of electrical conductivity on channel width is an experimental observation and does not require any assumptions about the boundary conditions. The saturation of conductance for $l_{\mathrm{ee}} > W$ highlights the importance of two characteristic length scales, $l_{\mathrm{ee}}$ and $W$, in determining the dominant scattering mechanism in the system. However, the plot of $G/G_{\mathrm{ball}}$ as a function of $l_{\mathrm{mfp}}/W$ (see Fig. \ref{fig:G-ballistic}) does not show a similar dependence, which rules out any interpretation of the data that relies on mean free path effects. In addition to these points, the simultaneous observation of WF-law violation, the decrease of $dV/dI$ with injected current density, and the quadratic dependence of conductivity on the channel width within our devices (as presented in Fig.~2 of the main text) strongly suggest that the experimental evidence in our devices can only be explained using a hydrodynamic description of electronic transport, with $l_\mathrm{ee}$ as the dominant lengthscale of interaction.

\section{Motivation for the proposed formalism}
\label{sec:motivation}

The theoretical model developed in this work is a phenomenological one, aimed at providing an alternative mechanism for explaining the device-dependent variability in transfer characteristics across multiple devices. Although the framework is heavily inspired by the Landauer-B\"uttiker formalism (LBF), which is used to describe ballistic systems, we must note that LBF describes conductance in the form of non-trivial transmission probabilities and can be extended to non-ballistic cases by appropriate corrections and assumptions to the transmission probabilities. Here, we consider an effective number of current-carrying modes $M_{\mathrm{eff}}$ and an effective transmission probability $T_{\mathrm{eff}}$ within this formalism to capture the hydrodynamic picture. This way of interpreting the physics is similar to the "Landauerization" of the Boltzmann equation presented in Ref.~\cite{eliminate-landauer-sharvin-resistance}, where the Boltzmann equation describing hydrodynamic electrons is reformulated to understand the Landauer-Sharvin resistance of an electronic channel. Moreover, LBF has demonstrated substantial success in explaining scenarios beyond strictly ballistic systems, including time-dependent transport \cite{LBF-photon-assisted, LBF-time-dependent}, non-linear transport \cite{LBF-nonlinear}, and transport in magnetic systems \cite{LBF-magnetic}, among others. In the present model, which is an attempt to understand the negative $R_\mathrm{4P}$ in our devices, the channel is assumed to host a homogeneous electron fluid with dynamic viscosity $\eta$ and a well-defined macroscopic conductance relation (i.e. $\sigma_V=(n^2 e^2 W^2)/12\eta$, here $n$ is the number density, $e$ is the electronic charge and $W$ is the width of the channel). In this scenario, LBF can be considered an efficient heuristic model for addressing the problem. Even with the crude approximations used, useful insight and numerical estimates can still be obtained.

\section{Derivation of the proposed formalism} \label{Derivation}
The phenomenological model used in the main text is motivated from the Landauer-B\"uttiker formalism (LBF) routinely used in multiprobe geometries. 
Consider a multiprobe geometry with $N$ leads. The conductance along the path from probe $q$ to probe $p$ ($G_\mathrm{pq}$) is defined as:
\begin{equation}
    G_\mathrm{pq} \equiv \frac{2e^2}{h}\bar{T}_\mathrm{p\leftarrow q}
    \label{eqn:conductance}
\end{equation}

where $e$ is the electronic charge, $h$ is the Planck's constant, and $\bar{T}$ is the total transmission coefficient given by $\bar{T} = \mathcal{M}T$ ($\mathcal{M}$ is the total number of conduction modes available for conduction and $T$ is the transmission probability). Note that $\mathcal{M}\ge0$ and $0\le T\le 1$, therefore $G\ge0$. Under the condition of small bias (i.e. difference between the chemical potentials of the current carrying probes $\Delta\mu\ll\varepsilon +(\text{few } k_\mathrm{B} \mathcal{T})$, here $\varepsilon$ is the energy range over which the transmission functions are nearly constant and $\mathcal{T}$ is the temperature), current at the $p^\mathrm{th}$ probe is given as \cite{datta1997electronic}:
\begin{equation}
    I_\mathrm{p} = \sum_\mathrm{q~=~1}^{\mathrm{N}} \left(G_\mathrm{qp}V_\mathrm{p} - G_\mathrm{pq}V_\mathrm{q}\right) \xrightarrow{G_\mathrm{pq}~=~G_\mathrm{qp}}\sum_\mathrm{q~=~1}^{\mathrm{N}} G_\mathrm{pq}\left(V_\mathrm{p}-V_\mathrm{q}\right) \label{eqn:LBF}
\end{equation}
Here, $N$ is the total number of probes in the device geometry. 

We assume a homogeneous viscous channel quantified by an effective conductance $G_\mathrm{V}$, while the coupling exhibited by the adjacent current and voltage probes is quantified by effective conductances $G_\mathrm{a},~G_{\tilde{\mathrm{a}}}$. Under these assumptions:

\begin{itemize}
    \item $G_{12} = G_{21} = G_\mathrm{a}$ (quantifies coupling between probes $I^+$ and $V^+$),
    \item $G_{34} = G_{43} = G_{\tilde{\mathrm{a}}}$ (quantifies coupling between probes $I^-$ and $V^-$),
    \item $G_{14} = G_{41} = G_{23} = G_{32} = G_{13} = G_{31} = G_{24} = G_{42} = G_\mathrm{V}$,
    \item $V_4 = 0$ (grounded)
\end{itemize}

Accordingly, we get the following expressions:
\begin{align}
I_1 = & (G_\mathrm{a} + 2G_\mathrm{V})V_1 - G_\mathrm{a} V_2 - G_\mathrm{V} V_3 = I^+\\
I_2 = & (G_\mathrm{a} + 2G_\mathrm{V})V_2 - G_\mathrm{a} V_1 - G_\mathrm{V} V_3 = 0 \label{I2}\\ 
I_3 = & (G_{\tilde{\mathrm{a}}}+2G_\mathrm{V})V_3 - G_\mathrm{V} V_1 -G_\mathrm{V} V_2 = 0 \label{I3}\\
I_4 = & -G_\mathrm{V} V_1 - G_\mathrm{V} V_2 - G_{\tilde{\mathrm{a}}} V_3 = I^-
\end{align}

The net current $I$ is given as:
\begin{equation}
    I = I^+ - I^- = I_1-I_4= (G_\mathrm{a} + 3G_\mathrm{V})V_1 - (G_\mathrm{a} - G_\mathrm{V}) V_2 - (G_\mathrm{V} - G_{\tilde{\mathrm{a}}})V_3
\end{equation}

Using Eq. \ref{I2} and \ref{I3}, we obtain the following relations:
\begin{itemize}
    \item $I_2 - I_3 = 0 = (G_\mathrm{V} - G_\mathrm{a})V_1 +(G_\mathrm{a}+3G_\mathrm{V})V_2 - (G_{\tilde{\mathrm{a}}}+3G_\mathrm{V})V_3$ \label{I2minusI3}
    \item $I_2 + I_3 = 0 = -(G_\mathrm{V} + G_\mathrm{a})V_1 +(G_\mathrm{a}+G_\mathrm{V})V_2 + (G_{\tilde{\mathrm{a}}}+G_\mathrm{V})V_3$ \label{I2plusI3}
\end{itemize}

These equations can be rewritten in matrix form as:
\begin{equation}
    {\begin{pmatrix}   G_\mathrm{a} - G_\mathrm{V}\\   G_\mathrm{a} + G_\mathrm{V}   \end{pmatrix}} V_1
    =
    {\begin{pmatrix}   G_\mathrm{a} + 3G_\mathrm{V} & -G_{\tilde{\mathrm{a}}}-3G_\mathrm{V}\\   G_\mathrm{a} + G_\mathrm{V} & G_{\tilde{\mathrm{a}}}+G_\mathrm{V} \end{pmatrix}}
    {\begin{pmatrix}   V_2\\  V_3   \end{pmatrix}} 
    \label{matrix-equation}
\end{equation}

We can now obtain expressions for $V_1$, $V_2$ and $V_3$ by solving the matrix equation as indicated above. The two-probe and four-probe resistances are respectively given as:  

\begin{align}
    R_\mathrm{2p} = & \frac{V_1 - V_4}{I_1-I_4}  = \frac{1}{8} \left(\frac{1}{G_\mathrm{V}}+\frac{1}{G_\mathrm{a}+G_\mathrm{V}}+\frac{1}{G_{\tilde{\mathrm{a}}}+G_\mathrm{V}}\right) \\
    R_\mathrm{4p} = & \frac{V_2 - V_3}{I_1-I_4} = \frac{1}{8G_\mathrm{V}} \frac{G_\mathrm{a} G_{\tilde{\mathrm{a}}} - G_\mathrm{V}^2}{(G_\mathrm{a} + G_\mathrm{V})(G_{\tilde{\mathrm{a}}}+G_\mathrm{V})} \label{R4p expression}
\end{align}

From the above equations, we see that $R_\mathrm{2p}$ cannot be negative, whereas the $R_\mathrm{4p}$ can be negative or positive depending on the sign of $G_\mathrm{a} G_{\tilde{\mathrm{a}}}-G_\mathrm{V}^2$. Therefore, it is the competition between the magnitudes of different conductances that can cause negative $R_\mathrm{4P}$. 

\begin{figure}[tbh]
    \centering
    \includegraphics[width=0.83\linewidth]{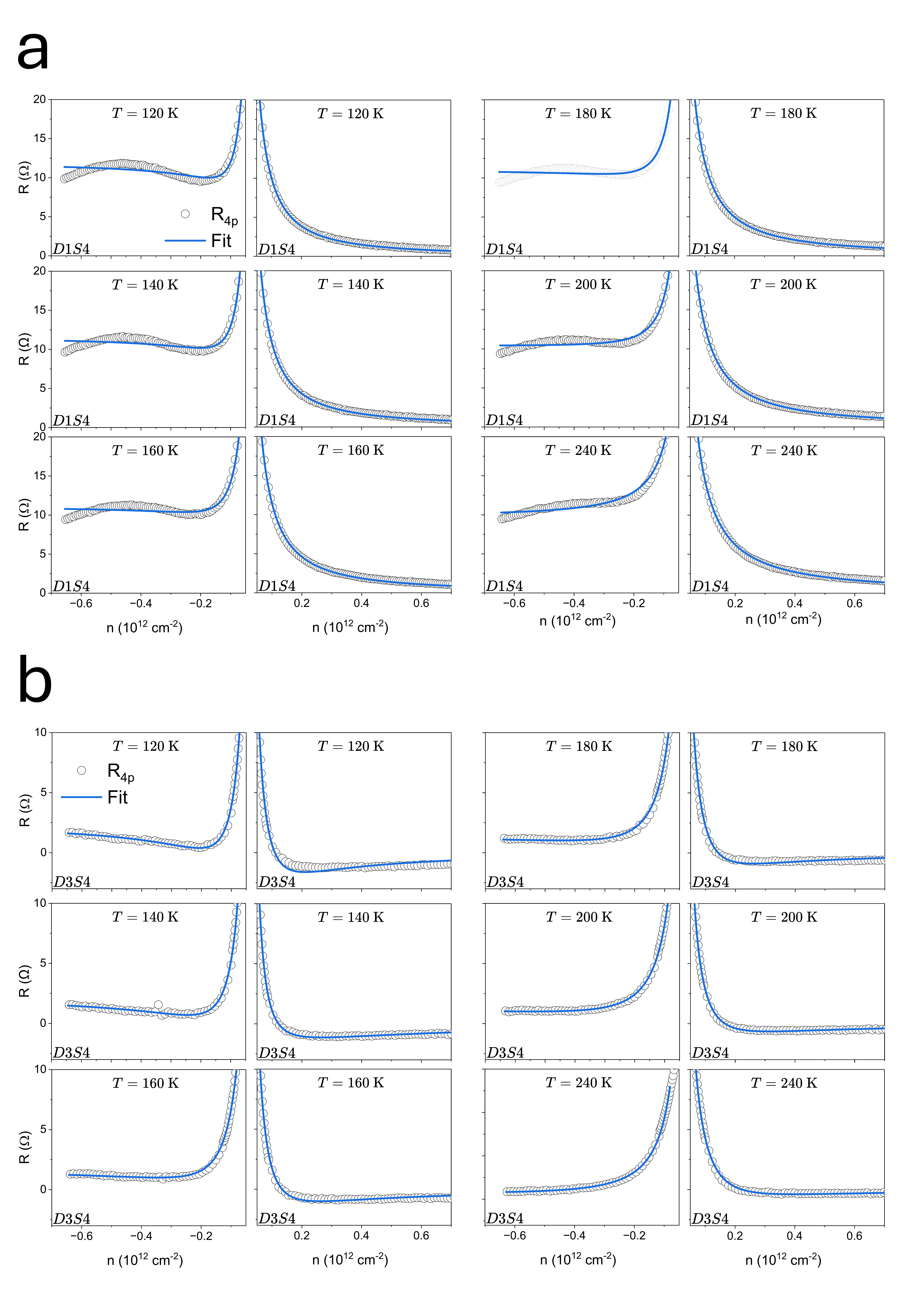}
    \caption[Quality of best fits]{\textbf{Quality of best fits:} $R_{\mathrm{4p}}$ vs $n$ for both electron and hole doped regimes ($\lvert n \rvert \ge 5 \times 10^{10}$~cm$^{-2}$) at different temperatures ($120$~K $\leq T \leq 240$~K) for devices (a) D1S4 and (b) D3S4. Along with the experimental data points (white circles), their best-fit curves have also been drawn, represented by solid blue lines.}
    \label{fig:best-fits}
\end{figure}

Eq.~(\ref{R4p expression}) is used to fit the four-probe resistance data. However, electrical transport at lower temperatures is affected by factors such as quantum interference effects \cite{Beenakker1991_book}, shot noise \cite{DanneauShotNoise2009}, etc. The formalism derived in Sec.~\ref{Derivation} is phenomenological in nature and does not take such effects into account. Therefore, it is not adequate to describe the electrical transport at low temperatures where such effects are prominent and impact the four-probe resistance on both electron and hole sides. Hence, our fitting equation remains valid for all electrical resistance data for temperatures above 100 K across multiple devices. The measured $R_\mathrm{4p}$ and the best fit for two of the devices used in this study is provided in Fig.~\ref{fig:best-fits}.

\section{Optimisation of the fitting scheme} \label{Fitting scheme}
In order to fit the measured resistance data $R_\mathrm{4p}(n)$, various electric conductance components are considered as $G_i = a_i n^j$ where $a_i$ and $j$ are fitting parameters. The parameter $j$ depends on the nature of the underlying scattering mechanism (tunnelling, ballistic, diffusive, hydrodynamic, etc.). We write:
\begin{gather}
    G_\mathrm{a} = a~n^b \\
    G_{\tilde{\mathrm{a}}} = a'~n^{b'}\\
    G_\mathrm{V} = c~n^{2-d}
\end{gather}
In certain devices (D1S6, D4S6), there exists a high asymmetry in the contact resistance at the metal-graphene interface, which is not accounted for in Eq. \ref{R4p expression}. This can be incorporated into the fit by introducing a device-specific constant, $R_{\mathrm{sym}}$, that accounts for the asymmetry between the contact resistances.
Therefore, the master fitting equation is given as:

\begin{equation} \label{master fitting equation}
    R_\mathrm{4p}(n) = R_{\mathrm{sym}} + \frac{|n|^{-2 + d} \left( a a' ~|n|^{b + b'} - c^2 ~|n|^{4 - 2d} \right)}{8c \left( a ~|n|^b + c ~|n|^{2 - d} \right) \cdot \left( a' ~|n|^{b'} + c ~|n|^{2 - d} \right)}
\end{equation}

\begin{figure}[tbh]
    \centering
    \includegraphics[width=0.8\linewidth]{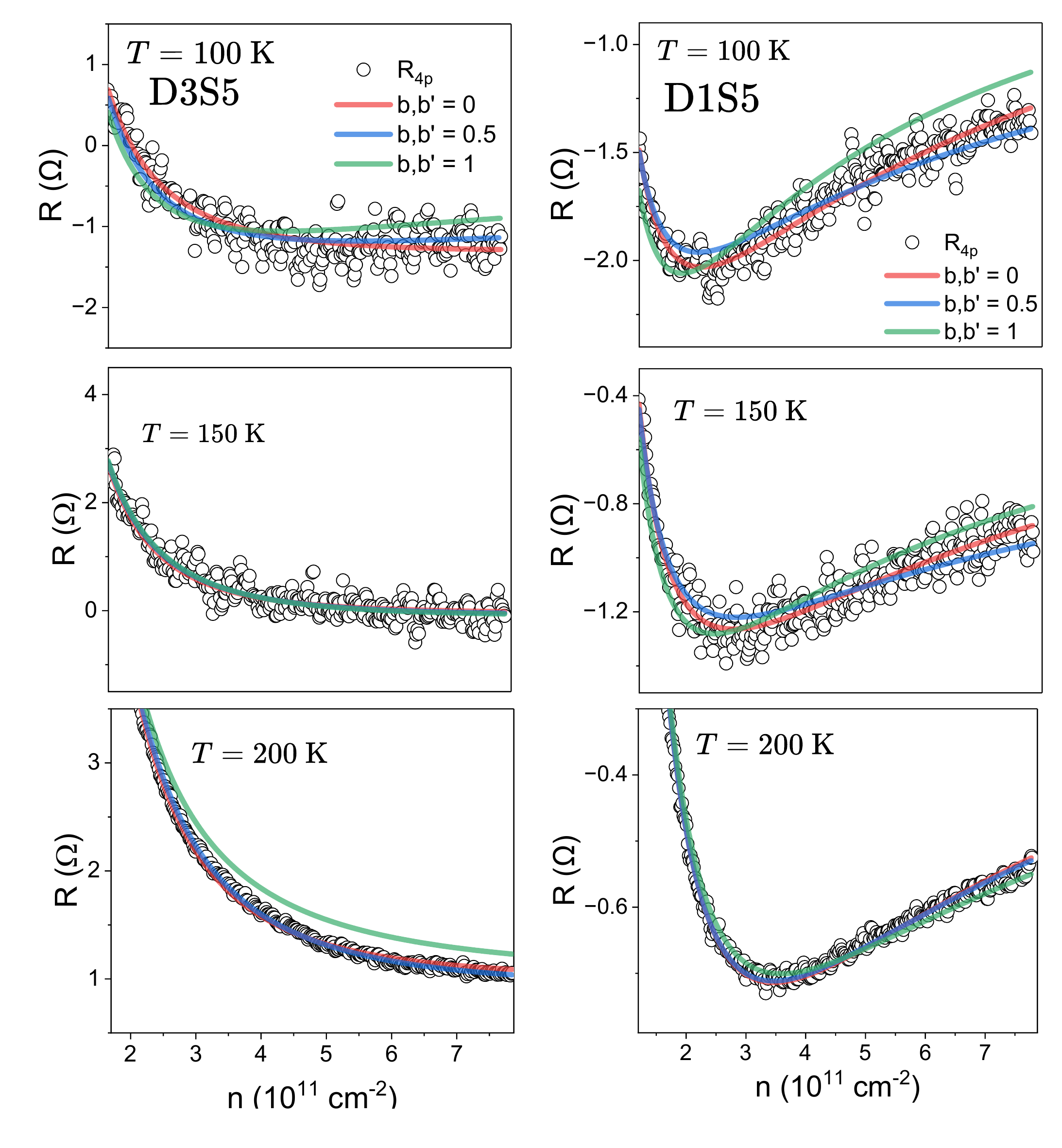}
    \caption[Comparison of the fitted curves]{\textbf{Comparison of the fitted curves}: $R_\mathrm{4p}$ vs $n$ for devices D3S5 (left) and D1S5 (right) evaluated at $T = 100~\mathrm{K},~150~\mathrm{K}~\text{and}~200~\mathrm{K}$ in the electron doped regime. The white circles indicate the experimentally measured resistance, whereas the solid lines indicate the best fitted curves for $b = b' = 0~(\text{red}),~0.5~(\text{blue})~\text{and}~1~(\text{green}).$}
    \label{fig:fit-comparison}
\end{figure}

The above equation can be further simplified by considering: 
\begin{itemize}
    \item Symmetry in the device geometry\\
    \item $R_{\mathrm{sym}} = 0$ (in high-quality devices)
\end{itemize}
The parameters $b$ and $b'$ are expected to be the same on the basis of the symmetric nature of the device geometry. Fittings with $b = b' = 0,~0.5,~\text{and}~1$ were performed (see Fig.~\ref{fig:fit-comparison}) with $b = b' = 0$ giving the best fit.

Using these simplifications, the final number of fitting parameters reduces to 4, namely $a,~c,~d$ and $a'$.

\begin{equation} \label{simplified R4p expression}
    R_\mathrm{4p}(n) = \frac{|n| ^{d-2} \left(a a'-c^2 ~|n| ^{4-2 d}\right)}{8 c \left(a+c~|n| ^{2-d}\right) \left(a' + c~|n| ^{2-d}\right)}
\end{equation}

Eq.~\ref{simplified R4p expression} has been used to fit the devices D1S4, D1S5, D3S4, D3S5 and D5S5. In order to fit devices D1S6 and D4S6, Eq. \ref{simplified R4p expression} was used in conjunction with the offset $R_{\mathrm{sym}}$.

\section{Extension of the proposed formalism }
The above derivation assumes that the channel is homogeneous and that an effective transmission coefficient, or equivalently, an effective conductance, can be defined. As such, the exact expression for the effective conductance may vary depending on the model. For example, if multiple experiments reveal that the channel is diffusive, Drude conductance may be chosen as the effective conductance. In our case, multiple experiments point towards a viscous channel (refer to Fig. 2 of the main text) and hence the effective conductance assumes a viscous form (refer to Sections \ref{sec:motivation}-\ref{Fitting scheme}). Hence, the model remains flexible across multiple cases and can be implemented in different geometries with appropriate modifications. 

\begin{figure}[tbh]
    \centering
    \includegraphics[width=1\linewidth]{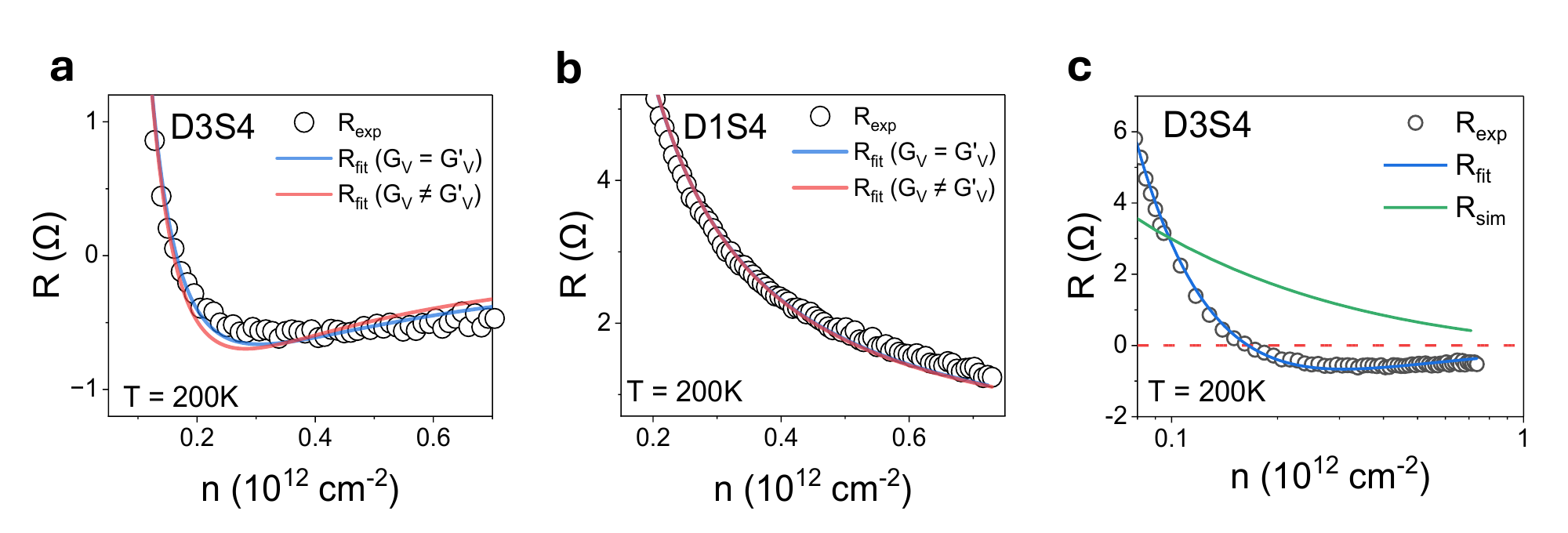} 
    \caption[Application of the proposed model described in Sections \ref{sec:motivation}-\ref{Fitting scheme} to experimental data]{\textbf{Application of the model described in Sections \ref{sec:motivation}-\ref{Fitting scheme} to experimental data:} (a) Fitted curve using the proposed model to experimentally determined $R_\mathrm{4p}$ vs $n$ using Eq. \ref{R4p expression} and Eq. \ref{R4p big expression} for a negative resistance device, corresponding to both the cases $G_\mathrm{V} = G'_\mathrm{V}$ and $G_\mathrm{V} \neq G'_\mathrm{V}$ respectively. (b) Same as (a) for a positive resistance device. (c) Comparison of experimental data with fit using the model described in Sections \ref{sec:motivation}-\ref{Fitting scheme} and numerical simulations. 
    \label{fig:fit_vs_experiment}}
\end{figure}

In this geometry, an improved version of Eq. \ref{R4p expression} where the straight viscous components ($G_{14}=G_{41}=G_{23}=G_{32}=G_\mathrm{V}$) and the cross viscous components ($G_{13}=G_{31}=G_{24}=G_{42}=G_\mathrm{V}'$) are taken to be different is given below:
\begin{equation}
R_{4\mathrm{P}}=\frac{G_{\mathrm{a}}\, G_{{\tilde{\mathrm{a}}}}-G_\mathrm{V}'^{\,2}}{2\,(G_\mathrm{V} + G_\mathrm{V}')\left[2 G_\mathrm{V} G_\mathrm{V}'+G_{{\tilde{\mathrm{a}}}}\,(G_\mathrm{V} + G_\mathrm{V}')+G_{\mathrm{a}}\,(2 G_{{\tilde{\mathrm{a}}}} + G_\mathrm{V} + G_\mathrm{V}')\right]}
\label{R4p big expression}
\end{equation}
However, we chose Eq. \ref{R4p expression} to fit the experimental $R_\mathrm{4P}$ data of our devices because of a lower number of parameters and producing a good quality of fits as Eq. \ref{R4p big expression}. Fig. \ref{fig:fit_vs_experiment} compares best fits to experimental data as obtained from Eq. \ref{R4p big expression} and Eq. \ref{R4p expression} along with $R_\mathrm{4p}$ evaluated from numerical simulations. The simulations do not match the experimental data well in the intermediate density regime. This can be attributed to numerical simulations performed using idealised hydrodynamic equations with idealised boundary conditions, with no information on the spatial distribution of impurity-induced potential fluctuations or other non-trivial couplings. As such, the viscous parameters obtained from numerical simulations assume their theoretical expressions, which may deviate in an actual experimental system; the same is accounted for in the proposed formalism, which treats the conductance as fitting parameters (\ref{Fitting scheme}) rather than theoretically derived quantities. However, the simulation data correctly captures the decreasing trend of $R_{\mathrm{4P}}$ with increasing number density.

\section{Competition between mean free path and momentum relaxation length}
The Drude mean free path contains contributions from all the electronic interactions in the system. However, the Drude mean free paths ($l_\mathrm{mfp}$) evaluated for our samples, as shown in Fig. 4b of the main text, are significantly larger than the electron-electron scattering length ($l_\mathrm{ee}$). This is in contrast to what is expected from Matthiessen’s rule, where scattering rates from independent mechanisms add up. This discrepancy can be explained by the fact that scattering mechanisms in a viscous electron fluid do not originate from independent sources. Previous theoretical works have also considered the addition of electrical conductances – the anti-Matthiessen’s rule \cite{guo2017higher, antimathessian}  in hydrodynamic systems – as an alternative, and transport experiments have validated these claims \cite{krishna2017superballistic}. 
Using a similar anti-Matthiessen approach, the Drude scattering time in our samples can be approximated as the sum of the different scattering mechanisms in the device. This gives $\tau_\mathrm{mfp}= \tau_\mathrm{mr}+ \tau_\mathrm{mc}$, where the momentum-conserving component is $\tau_\mathrm{mc}= \tau_\mathrm{ee}$ and the momentum-relaxing component is $\tau_\mathrm{mr}= \tau_\mathrm{e-imp}+ \tau_\mathrm{e-ph}$, considering impurity and phonon-mediated scattering. Since $\tau_\mathrm{ee} \ll \tau_\mathrm{mfp}$, we can say that the momentum relaxation length is given by $l_\mathrm{mr}=l_\mathrm{mfp}$ in devices of class II and III (where $l_\mathrm{mfp}\leq W$), and by $l_\mathrm{mr}=W$ in devices of class I (where $l_\mathrm{mfp}>W$).

\textbf{Dependence of conductivity on momentum relaxation length:}
The electrical conductivity of a rectangular two-dimensional graphene sheet of width \(W\) is given by \(\sigma \propto W^{2}/\eta\), where disorder or impurity scattering is negligibly small, the electronic momentum relaxes at the device boundary, and \(W \gg l_{\mathrm{ee}}\). In our category-II samples, the momentum–relaxation length \(l_{\mathrm{mr}}\) satisfies \(l_{\mathrm{ee}} \ll l_{\mathrm{mr}} \lesssim W\). We therefore approximate the hydrodynamic electrons as travelling across a channel of effective width \(l_{\mathrm{mr}}\), so that momentum relaxation occurs at the boundary of this redefined channel. With this approximation, we reuse the earlier expression as \(\sigma \propto l_{\mathrm{mr}}^{2}/\eta\).

\newpage
\bibliography{ref}
\newpage

